# Formation of giant planets around intermediate-mass stars

Heather F. Johnston,[1] O. Panić [1], B. Liu[2]★

[1] *School of Physics and Astronomy , University of Leeds, Woodhouse, Leeds LS2 9JT, England*
[2] *Astronomical Institute, School of Physics, Zhejiang University, 38 Zheda Road, Hangzhou, 310027 China*

**ABSTRACT**
To understand giant planet formation, we need to focus on host stars close to $M_\star$=1.7 M$_\odot$, where the occurrence rate of these planets is the highest. In this initial study, we carry out pebble-driven core accretion planet formation modelling to investigate the trends and optimal conditions for the formation of giant planets around host stars in the range of 1−2.4 M$_\odot$. We find that giant planets are more likely to form in systems with a larger initial disk radius; higher disk gas accretion rate; pebbles of ∼ millimeter in size; and birth location of the embryo at a moderate radial distance of ∼ 10 AU. We also conduct a population synthesis study of our model and find that the frequency of giant planets and super-Earths decreases with increasing stellar mass. This contrasts the observational peak at 1.7 M$_\odot$, stressing the need for strong assumptions on stellar mass dependencies in this range. Investigating the combined effect of stellar mass dependent disk masses, sizes, and lifetimes in the context of planet population synthesis studies is a promising avenue to alleviate this discrepancy. The hot-Jupiter occurrence rate in our models is ∼ 0.7−0.8% around 1 M$_\odot$ - similar to RV observations around Sun-like stars, but drastically decreases for higher mass stars.

## 1 INTRODUCTION

Giant planets play a decisive role in shaping the architecture of planetary systems. Studying the conditions and pathway for their formation is vital for our understanding on the dynamics and habitability of planetary systems, like our own solar system.

Nearly 10 − 20% of stars have been observed with giant planet(s) (Cumming et al. 2008; Johnson et al. 2010; Mayor et al. 2011), and this occurrence rate $\eta_J$ is found to increase with stellar metallicity (Gonzalez 1997; Santos et al. 2001, 2004; Fischer & Valenti 2005; Udry & Santos 2007; Johnson et al. 2010; Sousa et al. 2011; Mortier et al. 2012, 2013). These giant planets are typically observed at moderate orbital distances with a peak occurrence rate at around 2−3 AU (Suzuki et al. 2016; Fulton et al. 2021; Wolthoff et al. 2022).

Radial velocity (RV) observations inferred that $\eta_J$ increases with stellar mass from 0.5 to 2 $M_\odot$, ~15% around stars more massive than 1.5 $M_\odot$ while ~8% around sun-like GK stars (Johnson et al. 2010; Bowler et al. 2010; Borgniet et al. 2019). Meanwhile, Reffert et al. (2015) and Jones et al. (2015) pointed out that $\eta_J$ likely peaks at 1.5−2 $M_\odot$ and then readily drops for more massive stars. The giant planets are very rare around stars more massive than 2.5 $M_\odot$ ($\eta_J$~0). Recently, Wolthoff et al. (2022) combined data from three different RV surveys and reported that $\eta_J$ peaked around stars of $M_\star$≈1.7 $M_\odot$. Direct image observations also inferred that the intermediate-mass stars have the highest frequency of giant planets (Wagner et al. 2022). In addition, hot Jupiters might follow a similar stellar mass dependence as those orbiting at larger distances (Zhou et al. 2019; Gan et al. 2023).

We note that the above radial velocity measurements have difficulties in detecting planets around early-type main-sequence stars, due to the limited absorption lines and fast stellar rotation. Therefore, the above surveys search for (sub)giant stars that have evolved off their main-sequences, despite that the mass estimates for such "retired" stars are sometimes called into questions (Lloyd 2011; Schlaufman & Winn 2013). Asteroseismic studies of retired A stars (North et al. 2017; Stello et al. 2017) found no significant stellar mass derivations (up to 20%) compared to those presented in the aforementioned papers, dispelling the controversy around retired A-stars masses. Rotational velocity studies also confirmed that the masses of retired A stars have not been overestimated in previous analyses, supporting the reported trends with stellar mass (Ghezzi et al. 2018; Stock et al. 2018).

The exact origin of the above stellar mass dependent $\eta_J$ rate remains unclear. Stellar evolution - and the resultant impact on young protoplanetary disks - is thought to be a strong suspect. The evolutionary tracks of pre-main sequence stars with $M_\star$ > 2.5 M$_\odot$ differ significantly from their solar-mass analogs. Intermediate-mass stars become radiative early on in terms of disk lifetime (1 − 3 Myr), deviating from the well-established X-ray photoevaporation rate to far-ultraviolet (FUV) Kunitomo et al. (2021). However, when Nakatani et al. (2021) investigated the photoevaporation of grain-depleted protoplanetary disks around intermediate-mass stars, they found uncommonly low EUV and X-ray luminosities of intermediate-mass stars - compared to young, low-mass stars - due to absorption by the stellar atmosphere and loss of the host star's convective layer, mass-loss rates were diminished resulting in extended gas disk lifetimes around A-type stars. To retain coherence with the X-ray driven photoevaporation in our model, we will focus only on stellar masses < 2.5 M$_\odot$.

★ E-mail: bbliu@zju.edu.cn

Protoplanetary disk observations show strong correlations between the disk properties and the masses of their stellar hosts. (Sub)millimetre observations can probe total disk volume by measuring optically-thin dust continuum emission (Beckwith et al. 1990; Beckwith & Sargent 1991). Such studies conducted in several star-forming regions suggested that the total dusty disk mass increases with stellar mass, $M_{\rm dust} \propto M_\star^{1.3-1.9}$ (Pascucci et al. 2016; Stapper et al. 2022) and decreases with disk age (Andrews et al. 2009; Mann & Williams 2010; Mohanty et al. 2013; Barenfeld et al. 2016; Ansdell et al. 2016, 2017, 2018; Cieza et al. 2019). On the other hand, disk mass accretion rate exhibits a similar positive $M_\star$-dependency (Hartmann 1998; Muzerolle et al. 2005; Natta et al. 2006; Hartmann et al. 2006; Alcalá et al. 2014; Fairlamb et al. 2015; Manara et al. 2017). Furthermore, the stellar luminosity and mass correlation ($L_\star-M_\star$) also differs between low-mass T Tauri ($M_\star<2$ M$_\odot$) and Herbig Ae/Be stars ($M_\star>2$ M$_\odot$), due to the fact that more massive stars become more radiative dominated. This substantially influences the thermal evolution of protoplanetary disks (Booth et al. 2017; Miley et al. 2020).

Notably, the lifetime of protoplanetary disks might be considerably different among stars of various masses. This would profoundly impact the planet formation and migration processes (Burkert & Ida 2007; Alexander & Armitage 2009). Several mid-infrared (MIR) photometric studies from the Spitzer Space Telescope found that more massive stars ($M_\star>2$ M$_\odot$) deplete their disks more quickly, up to twice as fast as disks around low-mass stars (Carpenter et al. 2006; Dahm & Hillenbrand 2007; Kennedy & Kenyon 2009; Fang et al. 2012; Ribas et al. 2014, 2015). Such findings placed constraints on typical inner dust disk lifetimes at 3–5 Myr. On the other hand, long-lived disks are also found around intermediate-mass stars, known as Herbig Ae disks, with gas masses sufficient to foster giant planet formation at advanced ages of 5–10 Myr (Panić et al. 2008; Fairlamb et al. 2015; Kama et al. 2015; Ribas et al. 2018; Booth et al. 2019; Miley et al. 2019). Recent observations of debris disks, thought to be remnants of planet formation and devoid of gas, have found tens of disks with rich gas content such as CO, C I, and O I, at ages ≲50 Myr (Kóspál et al. 2013; Dent et al. 2014; White et al. 2016; Matrà et al. 2017; Marshall et al. 2017; Hughes et al. 2017, 2018; Higuchi et al. 2017, 2019). The reduced FUV photoevaporation model presented in Nakatani et al. (2021) discussed above may be the explanation for the primordial origin behind these aged gas-rich disks, again raising important implications for potential late-stage planet formation. The alternate hypothesis favors the secondary origin (shielded disk model), where an accumulation of C shields CO from photodissociation and thereby extends the disk lifetime (Moór et al. 2013, 2017, 2019; Kral et al. 2017, 2019; Hales et al. 2019; Marino et al. 2020). To summarize, the photoevaporation of these disks is largely driven by FUV and EUV, potentially challenging to quantify in theoretical models (Nakatani et al. 2021; Kunitomo et al. 2021).

A combination of the factors mentioned above may explain the peak and subsequent drop-off rate of giant planets with increasing $M_\star$. In fact, Liu et al. (2019) have conducted the state-of-the-art pebble-driven core accretion model to study the formation and migration of super-Earths and giant planets around stars of 0.1–1 M$_\odot$. They found that the most conducive conditions to form gas giants are systems having large initial disk size, high disk accretion rate and/or with stars of metal rich and high masses.

In this paper, we adapted the model to investigate giant planet formation around host stars in the stellar mass range of 1–2.4 M$_\odot$. We probe individual characteristics relevant for planet formation (*e.g.* initial disk radius (characteristic disk size) $R_{\rm d0}$, initial accretion rate $\dot{M}_{\rm g0}$, birth location of embryo $r_0$, birth time $t_0$) and investigate their stellar mass dependencies. The aim of this work is to act as a pathfinder to link observed trends to the physical assumptions in our model and establish reasons behind the distribution of the giant planet population, as inferred from Reffert et al. (2015), Jones et al. (2015), and Wolthoff et al. (2022). In order to do this, we seek to identify the key physical properties of protoplanetary disks that suit the giant planet formation, and discuss how these relate to stellar mass. This paper is organised as follows. The base of the model is described in Section 2. Section 3 discusses the illustrative simulations that visually present the function of the model and how parameters influence planetary growth and migration. The final planet mass and impact of birth conditions are explored in Section 4. The population synthesis study and conclusions are demonstrated in Sections 5 and 6, respectively.

## 2 METHOD

We employ the planet formation model of Liu et al. (2019) and recapitulate the key equations in this section. Readers are recommended to go through their Section 2 for details. The disk condition presented here is adapted to the circumstance around host stars of masses 1–2.4 M$_\odot$.

Due to the fact that the giant planets are overall uncommon, we optimise our disk model to suit the giant planets formation. Thus, these conditions will not necessarily correlate to average values observed in disks.

### 2.1 Disk model

We adopt a conventional $\alpha$ viscosity prescription for the gas disk angular momentum transport (Shakura & Sunyava 1973). We choose a viscously smooth disk without any structural discontinuities for simplicity. Note that the presence of gaps and rings in the disk caused by giant planet(s) or (magneto-)hydrodynamical instabilities can substantially impact the growth and migration of protoplanets. This is nevertheless beyond the scope of the current study. The disk accretion and surface density can be linked through $\dot{M}_{\rm g}=3\pi\nu\Sigma_{\rm g}$ with viscosity $\nu=\alpha_{\rm g}c_{\rm s}H_{\rm g}$, where $c_{\rm s}=H_{\rm g}\Omega_{\rm K}$ is the sound speed, $H_{\rm g}$ is the gas disk scale height and $\Omega_{\rm K}$ is the Keplerian frequency. The dimensionless viscous coefficient $\alpha_{\rm g}$ determines the global disk evolution and gas surface density. We set it to be a fix value of $10^{-2}$, the same as Liu et al. (2019).

We consider the disk angular momentum transport by gas accretion/spreading from the internal viscous stress (Lynden-Bell & Pringle 1974) and evaporation by high energy UV and X-ray photons emitted from the central star (Alexander et al. 2013). The evolution of the disk accretion rate is expressed as

$$\dot{M}_{\rm g} = \begin{cases} \dot{M}_{\rm g0}\left[1+\dfrac{t}{\tau_{\rm vis}}\right]^{-\gamma}, & t<t_{\rm pho}, \\ \dot{M}_{\rm g0}\left[1+\dfrac{t}{\tau_{\rm vis}}\right]^{-\gamma}\exp\left[-\dfrac{t-t_{\rm pho}}{\tau_{\rm pho}}\right], & t\geq t_{\rm pho}, \end{cases} \quad (1)$$

where $\dot{M}_{\rm g0}$ is the initial disk accretion rate; the $t-t_{\rm pho}$ term refers to the mass loss due to stellar photoevaporation; $\tau_{\rm vis}$ and $\tau_{\rm pho}$ are the characteristic viscous accretion and photoevaporation timescales from Eqs. (3) and (7) of Liu et al. (2019), $\gamma=(5/2+s)/(2+s)$, and $s$ is the gas surface density gradient.

The former term in the above equation represents the viscous driven disk evolution. The latter exponential term dictates that the stellar photoevaporation dominates the subsequent mass-loss at $t>t_{\rm pho}$ when $\dot{M}_{\rm g}<\dot{M}_{\rm pho}$. We specifically account for the X-ray driven photoevaporation, and the critical mass-loss rate is given by (Owen & Jackson 2012)

$$\frac{\dot{M}_{\rm pho}}{M_\odot \rm yr^{-1}} = 6\times 10^{-9} \left(\frac{M_\star}{1\,M_\odot}\right)^{-0.07} \left(\frac{L_X}{30\,\rm ergs^{-1}}\right)^{1.14}$$
$$= 3\times 10^{-8}\left(\frac{M_\star}{2.4\,M_\odot}\right)^{1.6}. \quad (2)$$

The latter equality in the above equation is derived from the stellar X-ray luminosity and mass relation $L_X \propto M_\star^{1.5}$ from Preibisch & Feigelson (2005).

It is worth mentioning that Kunitomo et al. (2021) examined the long-term disk evolution around stars of 0.5–5 $M_\odot$ and found that the X-ray luminosity of more massive stars decreases while the FUV luminosity rapidly increases (∼1 Myr for stars ≳3 $M_\odot$). The critical mass for this FUV-dominated photoevaporation mass-loss is thought to be around 2.5 $M_\odot$ when the Kelvin-Helmholtz timescale is comparable with the disk dispersal timescale (Kunitomo et al. 2021). Thus, the X-ray dominated photoevaporation regime that we investigate in this work with Equation 2 should be restricted to the lower stellar mass range of $M_\star<2.5\,M_\odot$ to retain coherence with the latest literature.

We employ a two-component disk structure, including an inner viscously heated region and an outer stellar irradiated region. The gas surface density and disk aspect ratio are adopted from Eqs. (8)-(13) of Liu et al. (2019) and [vis] refers to the inner viscously heated disk. [irr] refers to the outer disk region heated entirely by stellar irradiation (optically thin):

$$\frac{\Sigma_{\rm g}}{\rm g\,cm^{-2}} = \begin{cases} 480\left(\dfrac{\dot{M}_{\rm g}}{10^{-7}\,M_\odot\,{\rm yr^{-1}}}\right)^{1/2}\left(\dfrac{M_\star}{2.4\,M_\odot}\right)^{1/8} \\ \left(\dfrac{r}{1\,\rm AU}\right)^{-3/8} \quad [\rm vis], \\[6pt] 1440\left(\dfrac{\dot{M}_{\rm g}}{10^{-7}\,M_\odot\,{\rm yr^{-1}}}\right)\left(\dfrac{M_\star}{2.4\,M_\odot}\right)^{9/14} \\ \left(\dfrac{L_\star}{2.4^4\,L_\odot}\right)^{-2/7}\left(\dfrac{r}{1\,\rm AU}\right)^{-15/14} \quad [\rm irr], \end{cases} \quad (3)$$

and

$$h_{\rm g} = \begin{cases} 0.043\left(\dfrac{\dot{M}_{\rm g}}{10^{-7}\,M_\odot\,{\rm yr^{-1}}}\right)^{1/4}\left(\dfrac{M_\star}{2.4\,M_\odot}\right)^{-5/16} \\ \left(\dfrac{r}{1\,\rm AU}\right)^{-1/16} \quad [\rm vis], \\[6pt] 0.025\left(\dfrac{M_\star}{2.4\,M_\odot}\right)^{-4/7}\left(\dfrac{L_\star}{2.4^4\,L_\odot}\right)^{1/7} \\ \left(\dfrac{r}{1\,\rm AU}\right)^{2/7} \quad [\rm irr], \end{cases} \quad (4)$$

where $M_\star$, $L_\star$ are stellar mass and luminosity, and $r$ is the radial distance to the central star. The transition radius between these two disk regions is written as

$$r_{\rm tran} = 4.9\left(\frac{\dot{M}_{\rm g}}{10^{-7}\,M_\odot\,{\rm yr^{-1}}}\right)^{28/39}\left(\frac{M_\star}{2.4\,M_\odot}\right)^{29/39}$$
$$\left(\frac{L_\star}{2.4^4\,L_\odot}\right)^{-16/39}\,\rm AU. \quad (5)$$

The main disk solid reservoir - pebbles - are assumed to be constituted of 35% water ice and 65% silicate (Liu et al. 2019). The water-ice line $r_{\rm H_2O}$ is derived by equating the saturated pressure and $\rm H_2O$ vapor pressure (approximately in the disk radius at the temperature of 170 K). Additionally, in this paper we account for the sublimation of the silicate when the disk temperature is above 1500 K, which was neglected in Liu et al. (2019). The silicate sublimation line can be written as

$$\frac{r_{\rm Si}}{\rm AU} = \begin{cases} 0.91\left(\dfrac{\dot{M}_{\rm g}}{10^{-7}\,M_\odot\,{\rm yr^{-1}}}\right)^{4/9}\left(\dfrac{M_\star}{2.4\,M_\odot}\right)^{1/3} \quad [\rm vis], \\[6pt] 0.06\left(\dfrac{M_\star}{2.4\,M_\odot}\right)^{-1/3}\left(\dfrac{L_\star}{2.4^4\,L_\odot}\right)^{2/3} \quad [\rm irr]. \end{cases} \quad (6)$$

The silicate sublimation line is the maximum of these two values ($r_{\rm Si} = \max[r_{\rm Si,vis}, r_{\rm Si,irr}]$). When the pebbles drift inside of sublimation lines, their corresponding components rapidly evaporate as vapor and the pebble mass flux decreases accordingly. Since the pebbles only lose 35% of mass when they drift across $r_{\rm H_2O}$, their size only drops minorly. We neglect this r-dependent size variation of the pebbles in our model. We also do not implement any enhanced core accretion by icy pebbles that "stick" together more efficiently near the water-ice line (Okuzumi et al. 2012; Drazkowska & Alibert 2017; Hyodo et al. 2021)). In this paper, we do not account for any detailed grain growth processes. Instead, we assume that the dust has already fully grown and treat their final size to be a free parameter, indicative as either a fixed physical size or a fixed Stokes number. The above simplification ideally mimics that the grain growth is limited by the bouncing or fragmentation barriers (Güttler et al. 2010).

The inner edge of the disk is truncated by the stellar magnetospheric torque (Lin et al. 1996). We approximate the co-rotation cavity radius of the host star from Mulders et al. (2015b) as:

$$r_{\rm in} = \sqrt[3]{\frac{GM_\star}{\Omega_\star^2}} \simeq 0.06\left(\frac{M_\star}{2.4M_\odot}\right)^{1/3}\,\rm AU, \quad (7)$$

where $\Omega_\star$ is the stellar spin frequency.

To summarize, the inner edge of the disk in our model acts as the point where all planetary accretion and migration are terminated. This cutoff location from Mulders et al. (2015b) is in the range of ∼ 0.05–0.1 AU, depending on stellar mass, and is in agreement with the recent models from Flock et al. (2019) for FGK and M-type stars. However, Flock et al. (2016) conducted similar inner disk edge models for Herbig Ae stars, and the location was found to be stellar luminosity dependent, ranging from 0.09–0.42 AU. We assume $\Omega_\star$ is constant during the relatively short pre-main sequence stage for simplicity. While our $r_{\rm in}$ is static, we do adopt a dynamic silicate sublimation line $r_{\rm sil}$ in Equation 6, such that $r_{\rm sil}$ moves radially inward when the stellar luminosity decreases. As mentioned above, silicate sublimation occurs at disk temperatures above 1500 K thus no pebble accretion can proceed beyond this boundary as all refractory pebbles are evaporated.

### 2.2 Planet growth and migration

We consider the core mass growth of the protoplanetary embryo by pebble accretion. The embryo's starting mass is adopted as $10^{-2}\,M_\oplus$ if not otherwise stated. This assumption is commonplace throughout several planet formation studies, including Liu et al. (2019) and Voelkel et al. (2021). The solid accretion rate onto the planet's core reads

$$\dot{M}_{\rm PA} = \epsilon_{\rm PA}\dot{M}_{\rm peb} = \epsilon_{\rm PA}\xi\dot{M}_{\rm g}, \quad (8)$$

where $\epsilon_{PA}$ is the efficiency of pebble accretion, the formulas of which are adopted from Liu & Ormel (2018) and Ormel & Liu (2018) that include both 2D and 3D regimes and expressed as:

$$\epsilon_{PA} = \sqrt{\epsilon_{PA,3D}^2 + \epsilon_{PA,2D}^2}, \quad (9)$$

where $\epsilon_{PA,2D}$ and $\epsilon_{PA,3D}$ are defined from Liu & Ormel (2018) and Ormel & Liu (2018), respectively.

The pebble scale height is given by:

$$H_{peb} = \sqrt{\frac{\alpha_t}{\alpha_t + \tau_s}} H_g, \quad (10)$$

where $\tau_s$ is the Stokes number of the particles and $\alpha_t$ is the turbulent diffusion coefficient, approximately equivalent to the midplane turbulent viscosity when the disk is driven by magnetorotational instability (Johansen & Klahr 2005; Zhu et al. 2015; Yang et al. 2017).

Notably, $\alpha_t$ can principally differ from $\alpha_g$ - the average value of the global disk angular momentum transport efficiency - due to instances of layered accretion (Turner & Sano 2008). As noted in Liu et al. (2019), $\alpha_t$ is a more planet formation relevant parameter, responsible for dust stirring, pebble accretion as well as gap opening.

Whether pebble accretion is in 2D or 3D depends on the ratio between the radius of pebble accretion and the vertical layer of pebbles (Morbidelli et al. 2015). When $\tau_s < 10$, the majority of pebbles accrete onto the planet in a 'settling' fashion - where the particle is captured by the gravitational well of the planet and settles towards the planet (Huang & Ormel 2023). When $\tau_s > 10$, particles behave like a planet as they are experiencing less gas-aid. This means a 'ballistic'-style of accretion onto the planet - where the particle is not captured by the planet but ends up interior to the planet's orbit - becomes more prevalent with increasing $\tau_s$ (Huang & Ormel 2023). We thus ignore accretion in this regime and set $\epsilon_{PA}=0$.

Similar to Liu et al. (2019, 2020), we assume that the pebble and gas flux ratio remains a constant such that $\xi = \dot{M}_{peb}/\dot{M}_g$. Basically, when pebbles have a very low Stokes number, they are well-coupled to the disk gas. Therefore, these two drift inward at the same speed as the initial disk metallicity. When the pebbles' Stokes number is high, they drift faster than disk gas. In this case, in order to maintain a constant flux ratio, $\Sigma_{peb}/\Sigma_g$ becomes lower than the nominal disk metallicity (the pebble surface density gets reduced). We neglect the metallicity enrichment in the late rapid gas removal phase when the stellar photoevaporation dominates. In this paper a solar metallicity disk is assumed such that $\xi \equiv 0.01$.

A growing planet begins to clear the surrounding gas and opens a partial gaseous annular gap. As a consequence, the inward drift of pebbles stops at the outer edge of the planetary gap and the solid accretion terminates (Lambrechts et al. 2014). Such a planetary mass is referred to as the pebble isolation mass, and we adopt the formula from Bitsch et al. (2018) as

$$M_{iso} = 16 \left(\frac{h_g}{0.03}\right)^3 \left(\frac{M_\star}{3\,M_\odot}\right) \left[0.34\left(\frac{-3}{\log_{10}\alpha_t}\right)^4 + 0.66\right]$$
$$\left[1 - \frac{\partial \ln P/\partial \ln r + 2.5}{6}\right] M_\oplus. \quad (11)$$

For the adopted $\alpha_t = 10^{-4}$, the pebble isolation mass is close to $10-20\,M_\oplus$ around stars of $1-2.4\,M_\odot$. We nevertheless point out that all pebbles cannot be filtered by the planetary gap opening. A non-negligible fraction of pebble fragments at the above-mentioned planetary gap can still pass through and replenish the planetary envelope (Chen et al. 2020) as well as the inner disk region (Liu et al. 2022; Markus Stammler et al. 2023). We do not take that into account in this work.

We note that pebble isolation and gap-opening are the relevant processes (see discussions in Johansen et al. (2019)). The gap-opening mass is, for instance, defined as that the surface density at the planet's location is reduced to 50% of the unperturbed value (Kanagawa et al. 2015), while the pebble isolation mass refers to a minimum perturbation for generating a local pressure maximum, typically corresponding to a 20−30% level drop in the surface density (Bitsch et al. 2018). Following by Johansen et al. (2019), we also scale the gap opening mass as $M_{gap}=2.3 M_{iso}$.

When the low-mass planets still accrete pebbles, the solid materials entering the planetary atmosphere would generate sufficient heating to prevent the further contraction of the surrounding gas. Therefore, we only follow the gas accretion when the planet grows beyond $M_{iso}$. The gas accretion rate can be expressed as

$$\dot{M}_{p,g} = \min\left[\left(\frac{dM_{p,g}}{dt}\right)_{KH}, \left(\frac{dM_{p,g}}{dt}\right)_{Hill}, \dot{M}_g\right]. \quad (12)$$

We adopt the gas accretion rate based on Ikoma et al. (2000):

$$\left(\frac{dM_{p,g}}{dt}\right)_{KH} = 10^{-5}\left(\frac{M_p}{10\,M_\oplus}\right)^4 \left(\frac{\kappa_{env}}{1\,cm^2\,g^{-1}}\right)^{-1} M_\oplus yr^{-1} \quad (13)$$

where $\kappa_{env}$ is the planet's gas envelope opacity. We assume $\kappa_{env}=0.05\,cm^2\,g^{-1}$ and that it does not vary with metallicity, as in Liu et al. (2019).

Physically, the Kelvin-Helmholtz thermal contraction in Equation 13 promotes the envelope mass growth in the first place. As the planet grows, the accreted gas is further limited by the total amount entering the planetary Hill Sphere, adopted from Liu et al. (2019) below:

$$\left(\frac{dM_{p,g}}{dt}\right)_{Hill} = f_{acc}\nu_H R_H \Sigma_{Hill} = \frac{f_{acc}}{3\pi}\left(\frac{R_H}{H_g}\right)^2 \frac{\dot{M}_g}{\alpha_g}\frac{\Sigma_{gap}}{\Sigma_g}. \quad (14)$$

where $\nu_H = R_H \Omega_k$ is the Hill velocity; $R_H = (M_p/3\,M_\star)^{1/3}$ is the Hill radius of the planet; and $\Sigma_{Hill}$ is the gas surface density near the planet's Hill sphere. In this paper, we set $f_{acc}=0.5$ to be the gas fraction that can be accreted by the planet's Hill sphere. Once the planet becomes sufficiently massive, accretion is dictated by the global disk gas inflow across the planetary orbit.

We adopt a combined migration formula by incorporating both type I and type II regimes (Kanagawa et al. 2018):

$$\dot{r} = f_{tot}\left(\frac{M_p}{M_\star}\right)\left(\frac{\Sigma_g r^2}{M_\star}\right) h_g^{-2} v_K, \quad (15)$$

and the migration coefficient is given by

$$f_{tot} = f_I f_s + f_{II}(1-f_s) \frac{1}{\left(\frac{M_p}{M_{gap}}\right)^2}, \quad (16)$$

where $M_p$ is the mass of the planet, $v_K$ is the Keplerian velocity and $f_I$ and $f_{II}$ correspond to the type I and II migration prefactors, respectively. The type II migration coefficient $f_{II}=-1$ whereas the type I migration coefficient $f_I$ sets the direction and strength of the type I torque, determined by the disk thermal structure and local turbulent viscosity $\alpha_t$ (Paardekooper et al. 2011). We choose the smooth function $f_s^{-1}=1+(M_p/M_{gap})^4$. This ensures that $\dot{r} \simeq \dot{r}_I$ when $M_p \ll M_{gap}$ and $\dot{r} \simeq \dot{r}_I/(M_p/M_{gap})^2$ when $M_p \gg M_{gap}$ (Kanagawa et al. 2018).

## 2.3 Stellar mass dependence for disk parameters

Since the disk properties are correlated with stellar mass, the growth and migration of the planet is also expected to feature an $M_\star$-dependence. Here we summarize the key assumptions for the disk parameters and their stellar mass dependencies as follows.

(i) initial disk accretion rate $\dot{M}_{g0}$: A steeper than linear correlation between disk accretion rate and stellar mass is obtained in literature studies, mainly around low-mass T Tauri stars (Hartmann 1998; Muzerolle et al. 2005; Natta et al. 2006; Alcalá et al. 2014; Fairlamb et al. 2015). However, this correlation appears to break down in disks around more massive stars. For instance, Lopez et al. (2006) found that although Herbig Ae stars have generally higher accretion than solar-mass stars, no super strong accretors are detected among them (but also see Wichittanakom et al. (2020)). A double power-law profile of $\dot{M}_g - M_\star$ might be a realistic fit across the whole stellar mass regime, i.e., a steep component for low-mass T Tauri stars and a shallow one for high mass stars (Alcalá et al. 2017). In this work we assume $\dot{M}_{g0} \propto M_\star$ for the considered stellar mass regime. This results in the disk mass increasing linearly with stellar mass.

(ii) stellar luminosity $L_\star$: Stellar luminosity has a dependence of stellar mass on the pre-main sequence evolutionary tracks which differ in its trend on the Hertzprung-Russel diagram for stars below and above 1.5 $M_\odot$. For low-mass stars, $L_\star$ roughly increases with $M_\star^{1\sim 2}$ (Baraffe et al. 2002). However, higher mass AF type stars follow $L_\star \propto M_\star^{3.5\sim 4}$. Importantly, $L_\star$ decreases with age for low mass stars, but has an upward turn and remains high for higher mass stars, due to their transition to the radiative regime (Palla & Stahler 1993; Siess et al. 2000; Baraffe et al. 2002). Liu et al. (2019) adopted the power-law index of 2 for investigating planet growth around M dwarfs. Here $L_\star \propto M_\star^4$ is chosen since we focus on a higher stellar mass regime. For simplicity, we assume the stellar luminosity does not evolve within the disk lifetime. Nevertheless, Miley et al. (2020) included the stellar luminosity evolution and they found that temperature profiles of disks around low- and intermediate-mass stars begin to diverge at around 2 Myr. Disks around stars where $M_\star \geq 1.5\ M_\odot$ become warmer over time due to the increasing stellar luminosity, while disks around stars where $M_\star \leq 1.5\ M_\odot$ cool in temperature as the stellar luminosity decreases (Miley et al. 2020). Stellar luminosity and corresponding disk temperature have a profound impact on planet formation and migration. While the stellar luminosity in our model does not evolve, we adopt a two-component disk structure (inner regions are viscously heated, and outer regions are subjected to stellar irradiation in Equations 3 and 4). The transition radius (Equation 5) between these two regions relies on the luminosity of the host star and disk accretion rate. Consequently, both the water and silicate sublimation line (Equation 6) evolve with time, depending on the extent of radial heating within the disk.

(iii) initial disk size $R_{d0}$: Bate (2018) performed sophisticated radiation hydrodynamical simulations of star formation in clusters. Their results indicated that the sizes of the early protostellar disks are poorly (or at most weakly) dependent on $M_\star$. However, Stapper et al. (2022) found that disks around Herbig stars were generally both larger and more massive than disks around T Tauri stars, but with the caveat that the largest Herbig disk is of similar size to the largest T Tauri disk. They also speculate that these massive and large disks originate from initial disk mass (equivalently to assumed $\dot{M}_{g0}$ that increases with $M_\star$) and subsequent disk evolution. As such, we assume that $R_{d0}$ has no correlation with $M_\star$, at least for the explored stellar mass regime we consider in this work. Instead, we explore the role of disk size in giant planet formation in later sections and optimize our models accordingly.

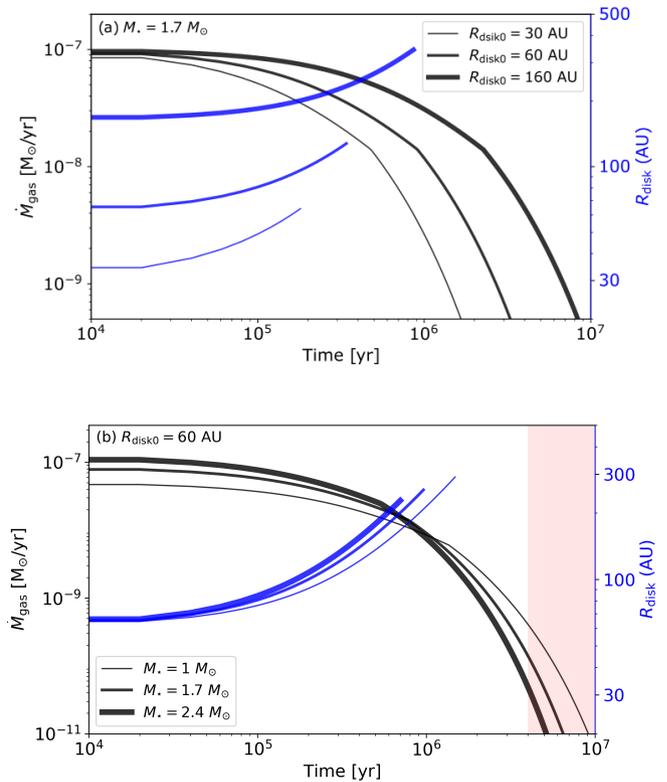

**Figure 1.** Evolution of disk accretion rate (black) and characteristic disk size (blue). At the early time, the disk angular momentum is governed by viscous accretion with the expanding size. Stellar x-ray evaporation dominates the gas depletion at the late time. The three different initial sizes of the disks are explored in panel (a), while the stars of three different stellar masses are considered in panel (b). The initial disk accretion rates are all adopted to be $10^{-7}\ M_\odot\ \mathrm{yr}^{-1}$ in panel (a), whereas $\dot{M}_{g0}$ follows $1.5 \times 10^{-7}(M_\star/2.4\ M_\odot)\ M_\odot\ \mathrm{yr}^{-1}$ in panel (b). The pink box indicates the regime where the 2.4 $M_\odot$ star is radiative (FUV-dominant photoevaporation).

## 2.4 Disk and pebble evolution

Based on the model setup described in previous subsections, here we demonstrate how the sizes and accretion rates of disks evolve with different initial conditions. Figure 1a shows the evolution of $\dot{M}_{gas}$ and $R_{disk}$ for three initial disk sizes $R_{d0}$ of 30 AU, 60 AU, and 160 AU around a central star of 1.7 $M_\odot$, whereas Figure 1b depicts the cases with the initial size of 60 AU around the stellar hosts of $M_\star=1\ M_\odot$, 1.7 $M_\odot$, and 2.4 $M_\odot$, respectively. Noticeably, the disk accretion rate (black) declines as the disk evolves viscously with an expanding radius (blue). A significant drop in disk accretion rate in the late stage is caused by the stellar photoevaporation when $\dot{M}_g$ is lower than $\dot{M}_{pho}$ (Equation 2 presides over Equation 1). This means that the X-ray photoevaporation regime is the dominant mass-loss mechanism in the system when $\dot{M}_g < \dot{M}_{pho}$. When $\dot{M}_{pho}$ is the dominant mass-loss mechanism, we cease tracking the expansion of the disk radius (blue lines), for simplicity due to the complexity involved in calculation radius expansion during this mass-loss regime.

In 1a) we see that smaller disks spread and their accretion rates decline much faster than larger disks. The lifetime of the disk is 1 Myr for the disk of $R_{d0}$=40 AU, while it is 4 Myr for the disk with four times larger initial size. On the other hand, the disks around higher mass stars evolve faster that those around lower mass stars, despite that disks are born with higher masses around the more massive stars (Figure 1b). This is because gas is depleted at an accelerated rate around more massive stars due to our assumptions on photoevaporation (Equation 2).

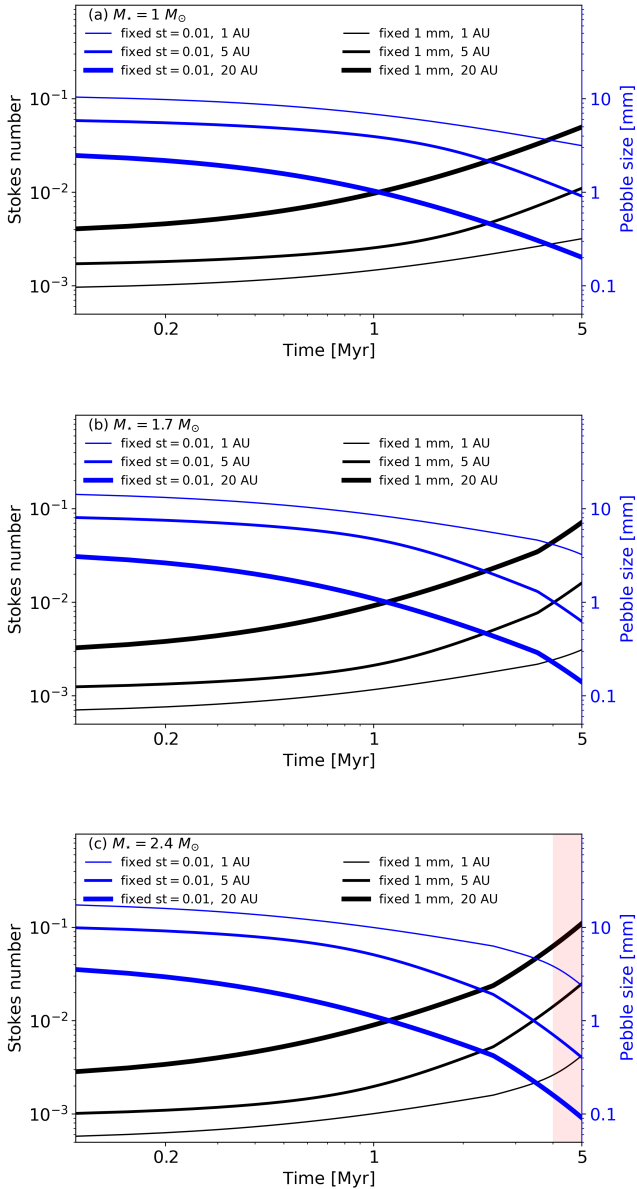

**Figure 2.** Evolution of pebbles at different time and radial distance for fixed physical sizes (black) and Stokes numbers (blue) in disks around stars of 1 $M_\odot$ (a), 1.7 $M_\odot$ (b), and 2.4 $M_\odot$ (c), respectively. The thickness of the line indicates the birth location of the embryo at 1 AU, 5 AU and 20 AU, respectively. Other parameters are $R_{d0}$=160 AU and $\dot{M}_{g0}=10^{-7}(M_\star/M_\odot)$ $M_\odot\,\mathrm{yr}^{-1}$. The pink box indicates the regime where the 2.4 $M_\odot$ star is radiative (FUV-dominant photoevaporation).

It is important to note that our model uses a solely X-ray driven photoevaporation rate (see Equation 2). We noted that Kunitomo et al. (2021) found that for stars with masses ≥ 2.5 $M_\odot$, their X-ray luminosity rapidly decreases ∼ 1 − 2 Myr and FUV becomes the dominating mechanism driving gas removal from the disk (see their Figure 11). This happens in response to the more massive stars becoming radiative. Thus, our adopted dominant X-ray photoevaporation regime is a fine assumption for our mass range of interest: 1−2.4 $M_\odot$ host stars. However, when $M_\star$=2.4 $M_\odot$, we do note that X-ray photoevaporation would only be the dominant gas removal mechanism for the first ∼ 4 Myr, at which point there would be a regime switch where FUV would become the prevalent mechanism - this is illustrated by the pink box in Figure 1 b. X-ray rates are well-understood and easily quantified, compared to FUV (Kunitomo et al. 2021). Hence, we do not know if it is more or less effective at clearing disk material than X-ray.

We account for the pebbles are either at a fixed size (1-mm) or at a fixed Stokes numbers. This 1-mm characteristic pebble size is derived from disk observations at mm-wavelengths measuring the spectral index (Draine 2006; Pérez et al. 2015). This is further supported by experiments conducted by Zsom et al. (2010), whom found that when colliding silicate dust, their growth is limited to mm-sizes due to the 'bouncing barrier'. We choose the fixed Stokes number to be $\tau_s$=0.01 as when $0.01 \leq \tau_s \leq 1$, the radial drift and pebble growth timescales are comparable (Birnstiel et al. 2012; Lambrechts & Johansen 2014; Johansen et al. 2019). Figure 2 demonstrates the time and radial distance dependencies of the pebbles' sizes in disks around stars of 1 $M_\odot$, 1.7 $M_\odot$, and 2.4 $M_\odot$. Again, the pink box in Figure 2 c denotes the regime switch where the 2.4 $M_\odot$ star would become radiative. The left-hand axis (black) shows how the particles' Stokes number evolves at various radial locations when the size is fixed at 1 mm. The typical Stokes number of millimeter-sized pebbles are closely in the range between $10^{-3}$ and 0.1. These pebbles have higher Stokes numbers at larger radial distances and later times. Their Stokes number would eventually exceed the unity as the gas disk further depletes by stellar photoevaporation.

The planet growth is crucially related to the Stokes number of the pebbles. The pebbles with a higher Stokes number ($\tau_s \gtrsim 1$) experience less gas-aid, and their aerodynamical behaviour is more analogous to that of a planetesimal. The accretion is less efficient accordingly (Ormel & Klahr 2010; Liu & Ormel 2018). In this regard, the planets cannot grow their masses substantially at very large orbital radii and/or late times when the pebbles' Stokes number is much higher than the unity.

Conversely, the right-hand axis (blue) of Figure 2 illustrates how the size of particles evolves over time when their Stokes number is fixed at 0.01 as at this value, the radial drift and pebble growth timescales are comparable (Birnstiel et al. 2012; Lambrechts & Johansen 2014; Johansen et al. 2019). Similarly, they have a larger physical size at closer-in orbits and earlier times. The dust particles have the size range between 100 $\mu$m and 1 cm, with slightly higher values in disks around more massive stars.

## 3 ILLUSTRATIVE SIMULATIONS

In order to identify the most favourable conditions for giant planet formation, we fix the model setups and only vary one individual parameter at a time ($R_{d0}$, $\dot{M}_{g0}$, $Z_d$, and $M_\star$). All pebbles are assumed to be 1-mm in size (see Section 2.4). Figure 3 shows the mass growth (upper panel) and orbital migration (lower panel) of a single embryo of $M_0=10^{-2}$ $M_\oplus$ (see Section 2.2) under the assumed disk and stellar conditions (Table 1). The shaded pink box in Figure 3d indicates the time at which a $M_\star$=2.4 $M_\odot$ star would decrease X-ray photoevaporation and instead FUV becomes the dominant gas removal mechanism as described in Kunitomo et al. (2021) (see Section 2.4).

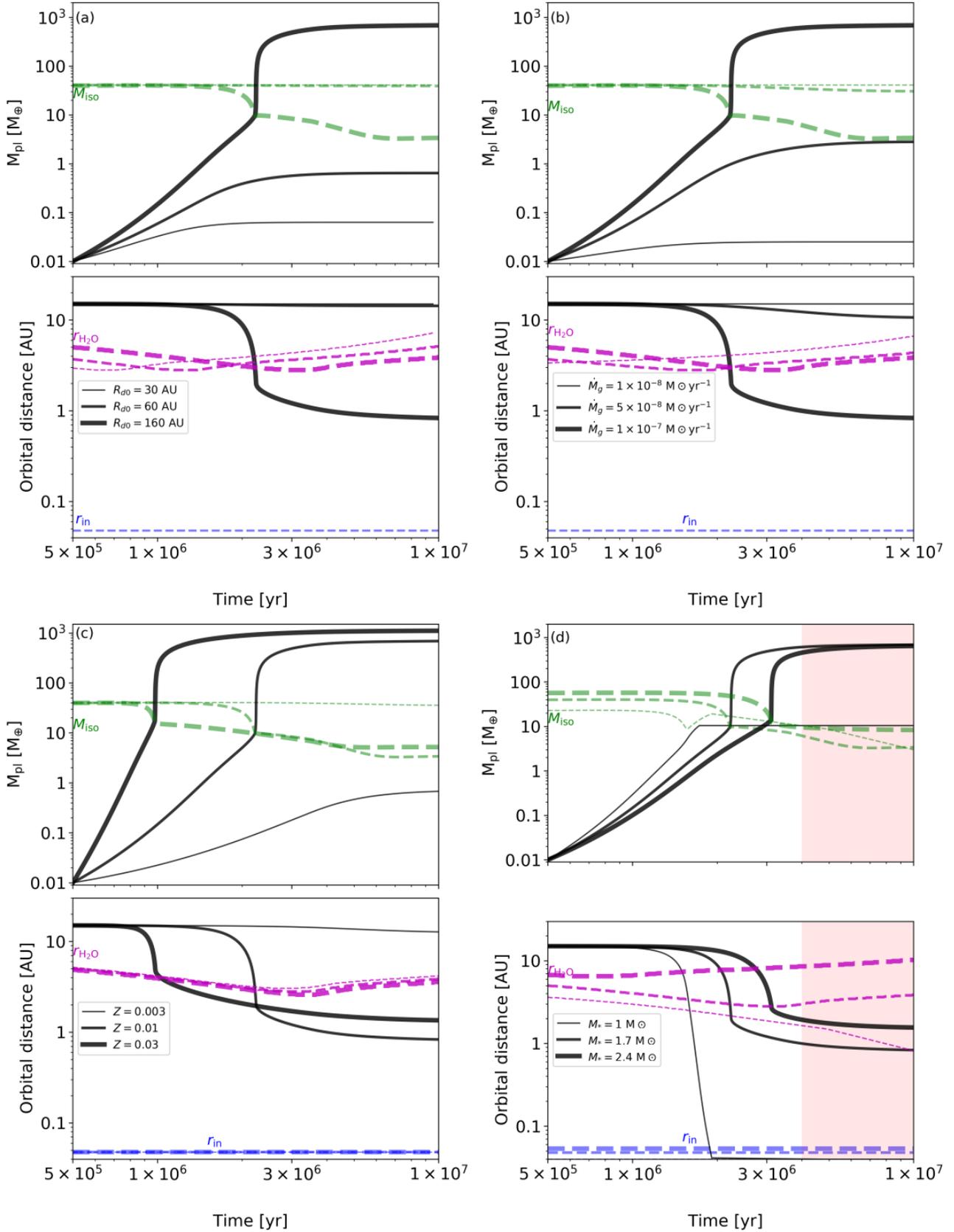

**Figure 3.** Planet mass growth (top) and orbital evolution (bottom) with the parameters adopted from Table 1: $R_{d0}$ (a); $\dot{M}_{g0}$ (b); $Z_d$ (c); and a fiducial model f1-f3 (d). The green line gives the isolation mass, while the magenta and blue lines represent the water ice line and inner disk edge (see Equation 7). Massive planets are more likely to form when the embryos are located $\geq 10$ AU in disks with larger sizes, high metallicity, and high accretion rates. The pink box indicates the regime where the 2.4 $M_\odot$ star is radiative (FUV-dominant photoevaporation).

| Name | $R_{d0}$ [AU] | $r_p$ [AU] | $\dot{M}_{g0}$ [$M_\odot$ yr$^{-1}$] | $Z_d$ | $M_\star$ [$M_\odot$] |
|---|---|---|---|---|---|
| S1 - S3 | 30, 60, 160 | 15 | $10^{-7}$ | 0.01 | 1.7 |
| D1 - D3 | 160 | 15 | $10^{-8}, 5\times 10^{-8}, 10^{-7}$ | 0.01 | 1.7 |
| Z1 - Z3 | 160 | 15 | $10^{-7}$ | 0.003, 0.01, 0.03 | 1.7 |
| f1 - f3 | 160 | 15 | $10^{-7}$ | 0.01 | 1, 1.7, 2.4 |

**Table 1.** Model parameters in Section 3.

### 3.1 Initial characteristic disk size

Figure 3a shows the results with disks of different initial characteristic disk sizes (S1 - S3 in Table 1). Stapper et al. (2022) examined ALMA archival data of 37 disks sround intermediate mass stars and found that such disks are generally larger than discs around lower mass stars (<1.5$M_\odot$ Ansdell et al. 2018), the largest disks in the Stapper et al. (2022) sample are 300-500 au in radius after about 5-7 Myr of viscous spreading, thus our $R_{d0}$ values were selected accordingly. In the disk of $R_{d0}$=30 AU, the embryo grows *in situ* to 0.06 $M_\oplus$. This is because of the vast majority of gas being depleted during the first 1 Myr (see Figure 1a). The embryo does not accrete a large amount of pebbles, and hence, is unable to migrate within the disk lifetime. In the disk of $R_{d0}$=60 AU, the planet grows at an increased rate and reaches 0.64 $M_\oplus$. However, it still does not grow massive enough to migrate significantly nor exceed the pebble isolation mass (green dashed line) to initiate rapid gas accretion.

In contrast to the smaller disks, the planet in the disk of $R_{d0}$=160 AU grows much faster. The rapid gas accretion proceeds further once the planet reaches $M_{iso}$ at $t\sim$2 Myr. The planet eventually grows into a giant planet with $M_p$=692 $M_\oplus$. The key factor that promotes the giant planet formation for the case of a larger initial disc size is that the disk accretion rates remain at high levels for a prolonged time when the sizes of their natal disks are larger, as demonstrated in Figure 1a. A high accretion rate fosters more massive planet growth as the solid mass accretion rate onto the planet is directly related to the gas accretion through Equation 8, thus a higher gas accretion rate results in a larger amount of solid material being accreted onto the planet.

### 3.2 Initial disk accretion rate

The growth of planets in disks with different initial disk accretion rates are shown in Figure 3b, (D1 - D3 in Table 1). Similar to the results of varying $R_{d0}$, more massive planets form in higher $\dot{M}_{g0}$ disks. Neither of the embryos becomes massive enough to exceed $M_{iso}$ when $\dot{M}_{g0}$=$10^{-8}$ or $\dot{M}_{g0}$=$5\times10^{-8}$ $M_\odot$yr$^{-1}$. Their mass growth is modest, reaching $M_p$=0.02 $M_\oplus$ and $M_p$=2.8 $M_\oplus$, respectively.

Yet the embryo in the disk of $\dot{M}_{g0}$=$10^{-7}M_\odot$yr$^{-1}$ has accreted the necessary amount of solid material to surpass $M_{iso}$ and underwent rapid gas accretion such that $M_p$=692 $M_\oplus$ (as mentioned in the above subsection). The key disk condition that drives giant planet formation is to sustain a high accretion rate during the planet growth period, as also described in Section 3.1.

### 3.3 Metallicity of the disk

We vary the metallicity of the disk - $Z_d$ scales with the pebble-to-gas flux ratio, *i.e.* a higher $Z_d$ means a larger pebble mass flux (Liu et al. 2019) - in Figure 3c (Z1 - Z3 in Table 1). More massive planets can form and form at a faster rate in a disk with a higher $Z_d$. The embryo is unable to accrete enough mass for even $M_p$=0.7 $M_\oplus$ when $Z_d$=0.001. When $Z_d$=0.01, the embryo is able to undergo runaway gas accretion when $t\sim$ 2 Myr and grows to $M_p$=547 $M_\oplus$.

Meanwhile, when $Z_d$=0.03, the embryo is able to grow massive $M_p$=1109 $M_\oplus$ in < 1 Myr. Disk metallicity increases the pebble accretion efficiency (see Equation 9), as such, the embryo is able to accumulate the majority of its solid content and undergo rapid gas accretion very early on. At $t \leq 2$ Myr, there would still be a large quota of gas content available and the accretion rate is still very high (as seen in Figure 1a). While high accretion rate remains the driving factor in massive planetary growth, disk metallicity can enhance this process by increasing the efficiency of pebble accretion, allowing the embryo to begin accreting gaseous material at earlier stages of disk evolution and for the giant planet to form faster.

### 3.4 Stellar mass

Finally, we examine and compare the planetary growth rates and orbital migration for a fiducial model of idealised giant planet formation around host stars of different masses (f1 - f3 in Table 1) in Figure 3d. Ergo: a large initial disk size; high accretion rate; and initial location at 15 AU.

The embryo around $M_\star$=1 $M_\odot$ grows to a moderate Super-Earth size: $M_p$=10 $M_\oplus$. The resultant planet is entirely rocky and does not surpass $M_{iso}$ to undergo rapid gas accretion. The growth of the embryo is stunted due to migrating past the edge of the inner disk in $\leq$ 2 Myr and is no longer able to accrete material beyond this boundary and is also not massive enough to trigger type II outward migration in the hope of accreting more mass.

Embryos around both $M_\star$=1.7 $M_\odot$ and 2.4 $M_\odot$ grow to be gas giants: $M_p$=692 $M_\oplus$ and $M_p$=655 $M_\oplus$, respectively (again, planet mass is calculated by our model for the specific combination of initial parameters). It should be noted that the embryo around the 2.4 $M_\odot$ grows and migrates at a notably slower pace than its counterparts around less-massive stars. This is because migration is less effective with increasing stellar mass, thus the extra timescale is needed to assist both growth and migration.

The faster/slower rate of planet growth depending on stellar mass is directly related to the pebble accretion efficiency (Equation 9). Liu & Ormel (2018) found that the pebble accretion efficiency decreases with increasing stellar mass. This is caused by a planet around a less massive star having a larger Hill's sphere and is able to accrete more pebbles (Liu & Ormel 2018). This results in the timescale for a planet to accrete enough solid material to surpass isolation mass and undergo gas accretion ($M_{iso}$) being longer around more massive stars (as seen in Figure 3 d when comparing $M_\star$=1.7 $M_\odot$ and $M_\star$=2.4 $M_\odot$).

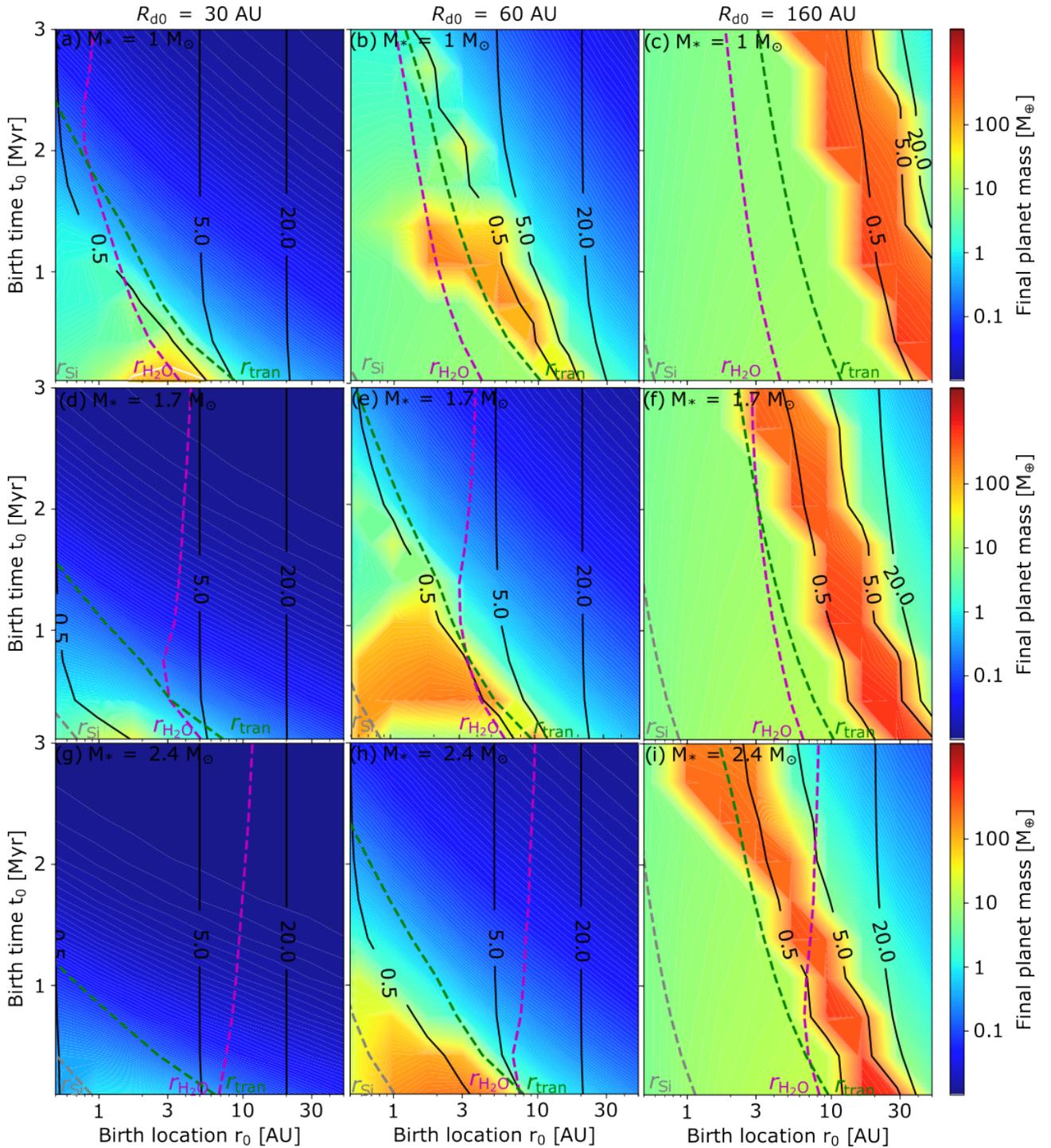

**Figure 4.** Maps for growth and migration of planets with a fixed particle size around stars of 1 $M_\odot$ (a, b, c), 1.7 $M_\odot$ (d, e, f), and 2.4 $M_\odot$ (g, h, i) with varying initial disk size $R_{d0}$=30 AU (1st column); $R_{d0}$=60 AU (2nd column); and $R_{d0}$=160 AU (3rd column), respectively. The black solid line refers to the final location of the planet, whereas the transition radius, silicate, and water sublimation line are marked as green, grey, and magenta dashed lines, respectively.

### 3.5 Exploring parameter space

In Sections 3.1 and 3.2, we found that both initial disk size and initial accretion rate play a vital role in dictating the amount of planetary growth of a single embryo. Now, we investigate this impact further by simulating planets at different birth times $t_0$ and locations $r_0$ around 1 $M_\odot$, 1.7 $M_\odot$, and 2.4 $M_\odot$ stars for varying disk sizes and accretion rates. The parameter setups S1 − S3 and D1 − D3 from Table 1 are adapted accordingly.

The motivation for such methods are while Figure 3 is sufficient in testing the impact of individual disk parameters on resultant giant planet formation and migration, we neglect both birth time and location of the embryo. Adopting this approach allows us to 'map out' the formation process and explore a wide range of birth times and locations for the embryo. Thus, we are able to keep an open mind as to where and when the giant planet formation process may begin.

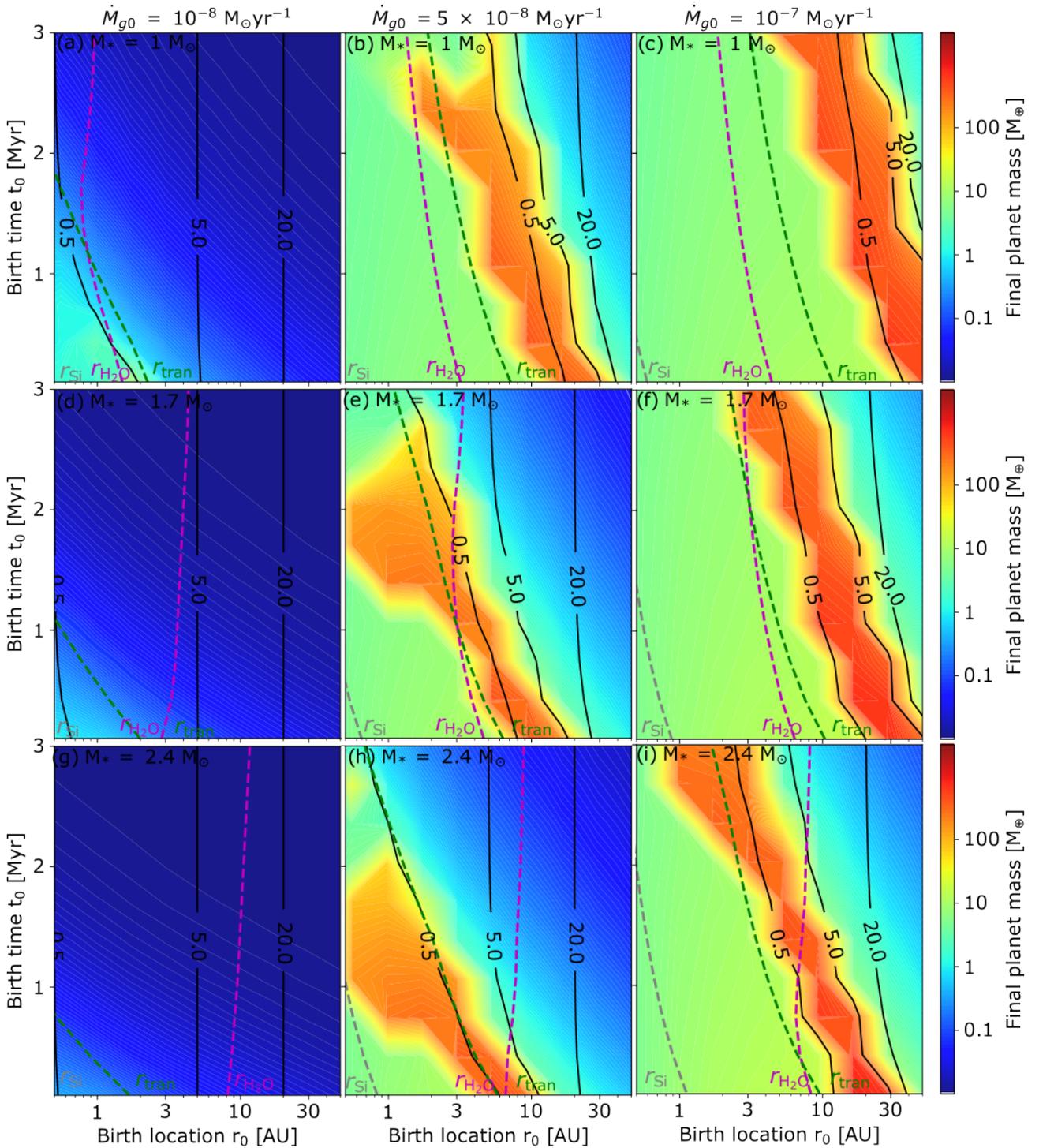

**Figure 5.** Maps for growth and migration of planets with a fixed particle size around stars of (a) 1 $M_\odot$, (b) 1.7 $M_\odot$, and (c) 2.4 $M_\odot$ with varying initial disk size $\dot{M}_{g0}=10^{-8}$ $M_\odot\,\mathrm{yr}^{-1}$ (1st column); $\dot{M}_{g0}=5\times 10^{-8}$ $M_\odot\,\mathrm{yr}^{-1}$ (2nd column); and $\dot{M}_{g0}=10^{-7}$ $M_\odot\,\mathrm{yr}^{-1}$ (3rd column), respectively. The black solid line refers to the final location of the planet, whereas the transition radius, silicate, and water sublimation line are marked as green, grey, and magenta dashed lines, respectively.

### 3.5.1 Varied disk size

We vary three initial characteristic disk size (30, 60 and 160 AU, respectively) and vary the central stellar mass to be 1 $M_\odot$, 1.7 $M_\odot$, and 2.4 $M_\odot$. Other parameters are the same as the fiducial run.

When $R_{d0}$=30 AU, we know from Figure 1a that around a 1.7 $M_\odot$ photoevaporation becomes the dominant gas removal mechanism in ≲ 2 Myr. Thus, it is to be expected that the outer disk regions at later times are dark blue in our maps (Figure 4 a, d, g). These denote failed embryos, where any kind of planet formation is impossible. In Figure 4c, no planet formation is possible beyond 1.5 Myr, even at close distances. Overall, a disk of size 30 AU is not conducive to any planet formation other than minor terrestrial growth to $M_p=1\,M_\oplus$ around a $1\,M_\odot$ star at very early time and $r_0 < 3$ AU.

In the 60 AU disk column, we find that the embryos grow and migrate modestly in Figure 4 b. This is because the gas in a relatively small disk gets quickly depleted within the first $\sim 3$ Myr (also see Figure 1a). Giant planets can only form at early phase in the disk region of < 10 AU. In the rest of the parameter space, the embryos still do not grow massive enough to migrate significantly nor exceed the pebble isolation mass to initiate rapid gas accretion. We also observe the shortened disk lifetime around more massive stars in Figure 4e and h, but to a lesser extend than when $R_{d0}$=30 AU around such stellar masses.

Disks evolve more slowly when their birth sizes are larger - as seen in Figure 1. In contrast to the smaller disks, the planets in the disk of $R_{d0}$=160 AU grow much faster and the giant planet formation zone gets more extended regardless of stellar mass (Figure 4 c, f, i). Therefore, the disk accretion rates remain at a high level for longer time when the sizes of their natal disks are larger, facilitating giant planet formation.

### 3.5.2 Varied disk accretion

In Figure 5, we assign three initial disk accretion rates from $\dot{M}_{g0}=10^{-8}$ (a, d, g), $\dot{M}_{g0}=5\times 10^{-8}$ (b, e, h), and $\dot{M}_{g0}=10^{-7}$ (c, f, i ). The disk accretion rate directly links to the gas mass of the disk. Thus, such an exploration can test the influence of the disk mass on final planet growth.

The corresponding migration maps are shown in Figure 5. Similar to the results of varying $R_{d0}$, more massive planets form in higher $\dot{M}_{g0}$ disks. None of the embryos becomes massive enough to exceed $M_{iso}$ when $\dot{M}_{g0}=10^{-8}$ (Figure 5 a, d, g). The mass growth is faster when $\dot{M}_{g0}=5\times 10^{-8}$ (b, e, h), but the less massive host star lends to a more successful giant planet forming zone - and more successful planet formation as a whole - especially the $1\,M_\odot$ case. Yet the embryos in the disk of $\dot{M}_{g0}=10^{-7}M_\odot yr^{-1}$ (c, f, i) have a higher chance to accrete sufficient solids to surpass $M_{iso}$ and underwent rapid gas accretion, regardless of stellar mass. The key disk condition that drives giant planet formation is to sustain a moderately high accretion rate during the planet growth period, as discussed in Section 3.2.

### 3.6 Summary of results

By examining Figure 3d, it is clear that it takes a significant amount of time for the protoplanetary embryo around a $2.4\,M_\odot$ with solar-$Z_d$ to accrete enough solid material to surpass the isolation mass and accrete gas. The embryo does not begin to accrete gas until $\sim 3$ Myr, *i.e.* prior to the FUV-dominated regime at 4 Myr (shaded pink box). This planet has completed its growth prior to the regime switch. If FUV is more effective at removing material than is the case for our assumed X-ray photoevaporation, the giant planet could be subjected to stripping of some of its gaseous content - a phenomenon that has been put forward to explain the hot super-Earth / Neptune desert (Lundkvist et al. 2016; Giacalone et al. 2022) . Conversely, if FUV is less effective at gas removal than the X-ray photoevaporation mechanism, it could potentially accumulate even more material to grow more massive than seen in our plots due to more material remaining in the environment and resulting in a prolonged disk lifetime.

The importance of having favourable disk conditions is further emphasised in Figures 4 and 5, which highlight the importance of having a large disk size and high accretion rate in producing giant planets by exploring the full parameter space. It is evident that planets are likely to grow more massive in disks with larger sizes (Figure 4) A similar trend is seen in Figure 5c where planets are most likely to grow massive in systems with a higher accretion rate (Figure 5). We can see that in a large disk size and high accretion environment, embryos are able to grow into gas giants from a range of birth times and locations. Despite giant planets being able to form as late as 3 Myr in Figures 4c (3rd column) and 5c (3rd column), it is possible that the FUV caveat above could impact any potential planetary growth when the embryo is formed at these later birth times.

In this circumstance, giant planets take $t \sim 3$ Myr to grow around $2.4\,M_\odot$ that they may never get the chance to grow massive. The gas removal mechanism switch between X-ray and FUV may be even more effective at disk clearing, leaving little material for any planetary growth.

## 4 GROWTH AND MIGRATION MAP

In this section we investigate the growth and migration of planets born at different birth times $t_0$ and locations $r_0$ around intermediate-mass stars of $M_\star=1\,M_\odot, 1.7\,M_\odot, 2.4\,M_\odot$, respectively. Since giant planets are rare, we optimise the parameters to promote giant planet formation and hence, our model does not necessarily correspond to 'average' disk properties but will still remain coherent with realistic values observed in disks. Thus, we adopt a relatively massive fiducial disk with $R_{d0}=160$ AU and $\dot{M}_{g0}=10^{-7}\,M_\star\,yr^{-1}$. We present the migration and growth maps by varying assumptions on pebble sizes, initial disk size, disk accretion rate, and birth mass of embryo in the following subsections. The parameter setups are listed in Table 2.

### 4.1 Fiducial: fixed pebble size

We assume all pebbles to have a constant size of 1 mm (derived from mm-wavelength observations of disks as described in Section 2.4 (Draine 2006; Pérez et al. 2015)) in the fiducial run of Table 2. In this regime, pebbles are all 1-mm but have varying Stokes number depending on their location in the disk and the stage of disk evolution, computed according to Equations 8-10. At later stages and greater disk radii, the pebbles have larger Stokes numbers resulting in them experiencing less gas coupling and exhibiting behaviour akin to that of planetesimals (see Figure 2). Such circumstances result in less efficient pebble accretion (Ormel & Klahr 2010; Ormel & Liu 2018).

The initial mass of the embryo is adopted as $10^{-2}\,M_\oplus$. The top panels of Figure 6 demonstrates the final mass and semimajor axis ($M_p, a_p$) the embryos eventually reach by initial $t_0$ and $r_0$, while the bottom panel provides the final water content in the corresponding planetary core. Colour gives the final planetary mass. The black lines describe the migration of growing planets during the first 3 Myr and the embedded values (0.5, 5, 20 AU) are the final locations of such planets. Some migrate within the first 3 Myr; while others take > 3Myr to reach their final location and do not reach their final location, as seen in Figure 6. The green, pink, and grey lines show the transition radius ($r_{tran}$), water ice line ($r_{H_2O}$), and silicate sublimation line ($r_{si}$).

We firstly examine the general planet formation trends around stars of $M_\star=1-2.4\,M_\odot$. In Figure 6, the embryos can grow into giant

MNRAS **000**, 1–20 (202)

| run | $R_{peb}$ [mm] | $\tau_s$ | $R_{d0}$ [AU] | $\dot{M}_{g0}$ [$M_\star yr^{-1}$] | $M_{p0}$ [$M_\oplus$] | $M_\star$ [$M_\odot$] |
|---|---|---|---|---|---|---|
| *run-fid* (fiducial) | 1 | $10^{-3}-10^{-2}$ | 160 | $10^{-7}$ | $10^{-2}$ | 1, 1.7, 2.4 |
| *run-tau* (Stokes number) | – | 0.1 | 160 | $10^{-7}$ | $10^{-2}$ | 1, 1.7, 2.4 |
| *run-emb* (embryo mass) | 1 | – | 160 | $10^{-7}$ | Eq. (10) | 1, 1.7, 2.4 |

**Table 2.** Model setups in Section 4.

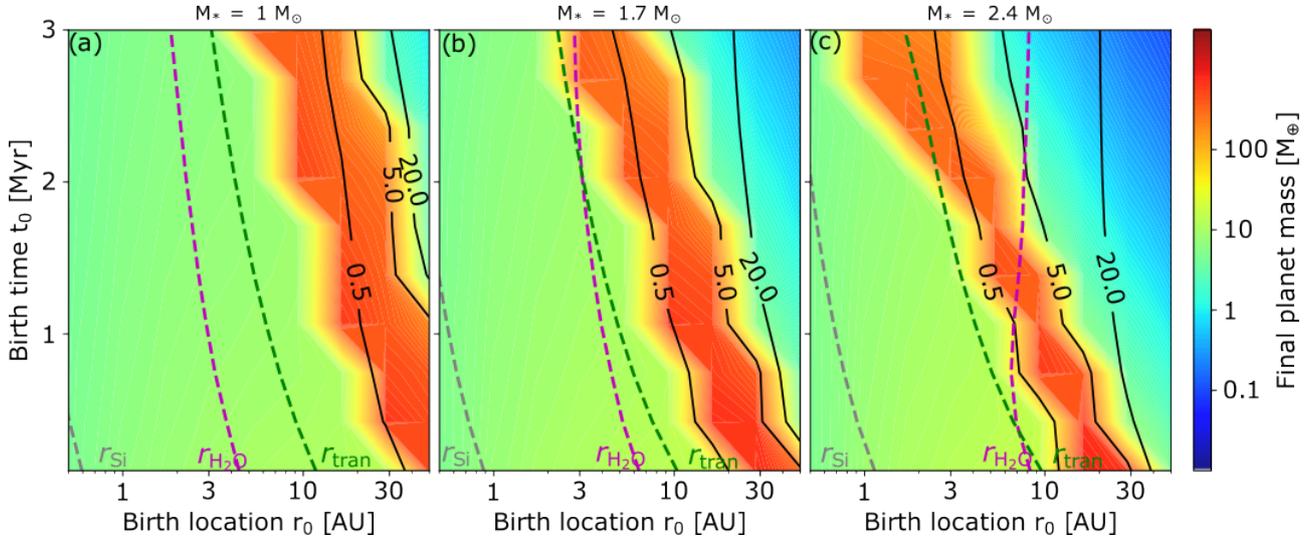

**Figure 6.** Maps for the growth and migration of planets as functions of initial time $t_0$ and $r_0$ around stars of (a) 1 $M_\odot$, (b) 1.7 $M_\odot$, and (c) 2.4 $M_\odot$. The black solid line refers to the final location of the planet, whereas the transition radius, silicate and water sublimation line are marked as green, grey and magenta dashed lines, respectively. The model parameters can be seen in *run-fid* of Table 2. Massive giant planets are favorable to form at $r_0 \sim r_{tran}$ and this formation zone becomes narrow as stellar mass increases.

planets when they are born between 20−30 AU in the early phase of $t_0 < 0.5$ Myr (red region) regardless of stellar mass. This giant planet forming zone becomes narrow and moves radially inward over time, spanning just a few AUs by $t = 3$ Myr. The most massive giant planets form when $t_0 \leq 1$ Myr, denoted by the dark red zone in the main red region. Thus, such embryos initially born at these locations would be able to grow into planets several times the mass of Jupiter. The rest of the giant planet region largely appears to grow planets between 100−400 $M_\oplus$ which is approximately the $M_{Saturn}$ to $M_{Jupiter}$. Contrarily, embryos grow to super-Earth planets (green region) when they are initially located interior to this giant planet forming region. This is due to their fast inward migration (Johansen et al. 2019; Liu et al. 2019). Such embryos accrete a modest amount of solids but are unable to achieve $M_{iso}$ and attain gas content before reaching the edge of the inner disk ($r_{in}$) where any accretion and migration ceases, as described in Equation 7 (Lin et al. 1996; Mulders et al. 2015b). Since the planets found in this region grow to $M_p \leq 10\,M_\oplus$, they are not massive enough to undergo type II migration as expressed in Equation 15 (Kanagawa et al. 2018). Thus, such planets cannot grow or migrate further once they reach $r_{in}$. Meanwhile, embryos with $r_0 \geq 40$ AU and $t_0 > 2$ Myr (blue region) accrete very little solids and remain low-mass because of inefficient pebble accretion at further out disk radii (Ormel & Liu 2018, blue region). Therefore, only the embryos born at moderate $r_0$ can facilitate the formation of massive giant planets. While we do not apply any optimised conditions to the snowline - *e.g.* icy grains, like those found at the water snowline, are more likely to stick together and form larger pebbles, increasing the efficiency of pebble accretion - we do not find that the water snowline fosters efficient planet formation in our models. In fact, the water snowline often falls outside this ideal region (red band). It is evident that the evolution of the disk combined with optimised initial conditions preside over the concept of one ideal location at forming giant planets.

*4.1.1 Planetary migration*

Now we compare how the growth and migration of the planets relate to stellar mass. For planets with specific mass and radial distance, the migration and pebble accretion can be written as follows

$$\tau_{mig} = f_{tot}^{-1} \left(\frac{M_\star}{M_p}\right) \left(\frac{M_\star}{\Sigma_g r^2}\right) h_g^2 v_K^{-1} \propto M_\star, \quad (17)$$

$$\tau_{grow} = \frac{M_p}{\dot{M}_{PA}} = \frac{M_p}{\dot{M}_{peb}\epsilon_{PA,2D}} \propto M_\star^{-1/3}, \quad (18)$$

where $\dot{M}_{peb} \propto \dot{M}_g \propto M_\star$, and $\epsilon_{PA,2D} \propto (M_p/M_\star)^{2/3}$ is adopted from the 2D shear regime (Liu & Ormel 2018). Simply, the migration timescale increases with $M_\star$ while the growth timescale decreases with $M_\star$. So although growth is more efficient around more massive stars, it cannot compete with the lengthened migration timescale. This relates back to our findings in Figure 3d where we find that the embryo that migrates the most efficiently is found around $M_\star = 1.7$ $M_\odot$ (growth is more efficient than in the 2.4 $M_\odot$ case and migration is less efficient than in the 1 $M_\odot$ case which migrates beyond the edge of the inner disk too quickly). While embryos remain fairly static in terms of migration when undergoing initial core accretion until $M_{iso}$ can be achieved, gas accretion is always accompanied by inward migration that is most efficient at small $M_\star$.

We find that the giant planet formation zone moves towards smaller $r_0$ and becomes shallower with $t_0$ as $M_\star$ increases (Figure 6b and c). As can be seen in Equation (17), this is because the planet migration rate declines with stellar mass, as discussed above. The planets in disks around more massive stars take longer time to enter into the inner disk.

Firstly, in the 1 $M_\odot$ case, embryos initially located close to the star are unable to become giant planets because they migrate too quickly, whereas embryos initially located far away from the star remain small because they grow too slowly. Meanwhile around more massive stars, the marginally enhanced growth rate of planets cannot compete with the slow rate of migration. This causes the giant planet forming region to be shifted closer to the star because such planets are able to sustain their growth without suffering from runaway migration and reaching the inner edge of the disk too quickly. The original green region at $t=0$ yr in the left side of Figure 6a ($r_0 \sim 15$ AU) now becomes red in Figure 6c as the giant planet forming region is radially limited and begins at a closer AU to the host star to begin with. Since the rate of migration is slower when 2.4 $M_\odot$ star, such embryos take a lot longer to reach $r_{in}$ than the same embryos found around $M_\star=1$ $M_\odot$ would, as evidenced in Figure 3 d. This prolongs the timescale for planet accretion thus allowing them to grow into more massive planets than those seen in Figure 6a. Similarly, in Figure 6c the embryos at $r_0>30$ AU undergo moderate migration and stall beyond 20 AU. As discussed in Section 2.4, in a fixed 1 mm regime, pebbles at large distances and later times have a higher Stokes number $\sim 0.1$. Such pebbles experience less gas-aid and behave in a planetesimal-like fashion which causes their accretion to be less efficient, resulting in inefficient planetary growth (Ormel & Klahr 2010; Ormel & Liu 2018). Hence, pebble accretion is limited at such large orbital distances. As a result, these embryos fail to grow into giant planets, which differs from the outcome in Figure 6a.

The giant planet formation window is also radially limited at later times for disks around more massive stars due to our assumption that the disk lifetime is shorter around more massive stars (as seen in Figure 1b), resulting in any late-stage giant planet formation in such systems becoming even more challenging. For instance, at $t_0=3$ Myr only embryos born at $r_0<5$ AU have a chance to grow sufficient massive around stars of $M_\star=2.4$ $M_\odot$. However, in Figure 6a embryos formed at $r_0\sim10-25$ AU can turn into giant planets when $t_0=3$ Myr. Hence, the giant planet forming region is notably smaller around a 2.4 $M_\odot$ star.

We note that although the pebble flux is higher in disks around more massive stars, the relative pebble-planet velocity also drops as $M_\star$ increases in the 2D shear regime. Thus, the overall pebble accretion rate for an embryo of initial $10^{-2}$ $M_\oplus$ is only modestly dependent on $M_\star$. Hence, the main discrepancy for planet formation around different mass stars is caused by $M_\star$-dependent migration.

### 4.2 Fixed Stokes number

In *run-tau* of Table 2, we explore the planet growth by assuming pebbles with a fixed Stokes number of $\tau_s=0.01$ rather than fixed size. $\tau_s=0.01$ is chosen as the timescales for radial drift and pebble growth are comparable, as described in Section 2.4 (Birnstiel et al. 2010; Lambrechts & Johansen 2014; Johansen et al. 2019). When the pebbles are in a fixed-Stokes system, they have varying sizes depending on their location in the disk and the stage of disk evolution - e.g. at later stages and further disk radii, the pebbles are smaller in size (see Figure 2). Such circumstances result in the $\mu$m-sized pebbles being too small to sustain substantial planetary growth. The other parameters are detailed in Table 2. We are primarily interested in how the variation of the pebbles' size assumption would impact the pattern of planetary migration and growth. Our *run-fid* with fixed 1−mm pebbles corresponds to a Stokes number between $10^{-3}-10^{-2}$, thus we can compare Figure 6 and 7 directly in the sense of comparing a low Stokes regime to a high Stokes regime.

The overall features are similar between Figure 7 and Figure 6, but in *run-tau* the giant planet formation zone is narrower compared to *run-fid* yet less horizontally inclined. Comparing Figures 6a and 7a at $t=0$ Myr, it is already evident that the giant planet forming region is less broad in *run-tau* than *run-fid* when $t_0 \geq 1.5$ Myr. In Figure 7b, the giant planet forming region spans a similar radial region of $20-50$ AU at $t_0=0$ Myr but quickly becomes narrower as birth time increases. By $t=3$ Myr, the outer region does not grow beyond the initial mass of the embryo between $r_0=20-50$ AU (dark blue) compared to Figure 6b where embryos located at the same radial distance are growing to $M_p \sim 1$ $M_\oplus$. Around a 2.4 $M_\odot$ in Figure 7c, giant planets only from between $20-40$ AU at $t=0$ Myr. This zone remains radially narrower than the fixed pebble size regime in Figure 6 as time progresses and less massive overall ($M_p \sim 100$ $M_\oplus$). The entire giant-forming feature is limited radially in a fixed Stokes regime for all stellar masses.

This can be understood by looking at Figure 2 that pebbles of 1 mm in size have the corresponding Stokes numbers of $\sim 10^{-3}-10^{-2}$. The pebbles with a higher Stokes number drift faster in the disk, and therefore, the embryos accrete in a less efficient 2D mode, as described in Equations 8 and 9 (Liu & Ormel 2018; Ormel & Liu 2018; Liu et al. 2020). Fast-drifting pebbles also mean they will be depleted more rapidly compared to that of the gas. This factor is not considered in our study, which may further limit the formation window of massive planets. In sum, the growth of the massive planets becomes more difficult when the pebbles have a large Stokes number.

It is also worth noting that we also conduct separate tests for pebbles at a fixed Stokes number of 0.01 and $10^{-3}$. We find that the resultant migration map for a fixed Stokes number of 0.01 is more resemble to the case when pebbles are fixed with millimeter in size. Nevertheless, for even lower Stokes number of $10^{-3}$, the planet growth turns less efficient again, mainly due to the fact that the accretion largely settle into the slow 3D accretion regimes. Overall, in our study the optimal size for planet growth is 1 mm, or approximately the Stokes number of $\sim 0.01$.

### 4.3 Varied embryo mass

The starting mass of the embryo is fixed to $10^{-2}$ $M_\oplus$ in previous sections. Streaming instability is a powerful mechanism that forms planetsimals from the collapse of many mm-sized pebbles (Youdin & Goodman 2005). It is a method that is particularly adept at forming planetesimals (Johansen et al. 2007, 2012; Bai & Stone 2010; Simon et al. 2016; Schäfer et al. 2017; Abod et al. 2019; Li et al. 2019). Liu et al. (2020) summarized from the literature streaming instability planetesimal formation simulations that the birth masses of the embryos are dependent on their evolutionary time and disk locations. Consequently, we examine the migration map by assuming varied initial embryos' masses according to Eq. (4) of Liu et al. (2020):

$$\frac{M_{p0}}{M_\oplus} = 6 \times 10^{-2} \, (\gamma\pi)^{1.5} \left(\frac{h_g}{0.05}\right)^3 \left(\frac{M_\star}{2.4 \, M_\odot}\right). \quad (19)$$

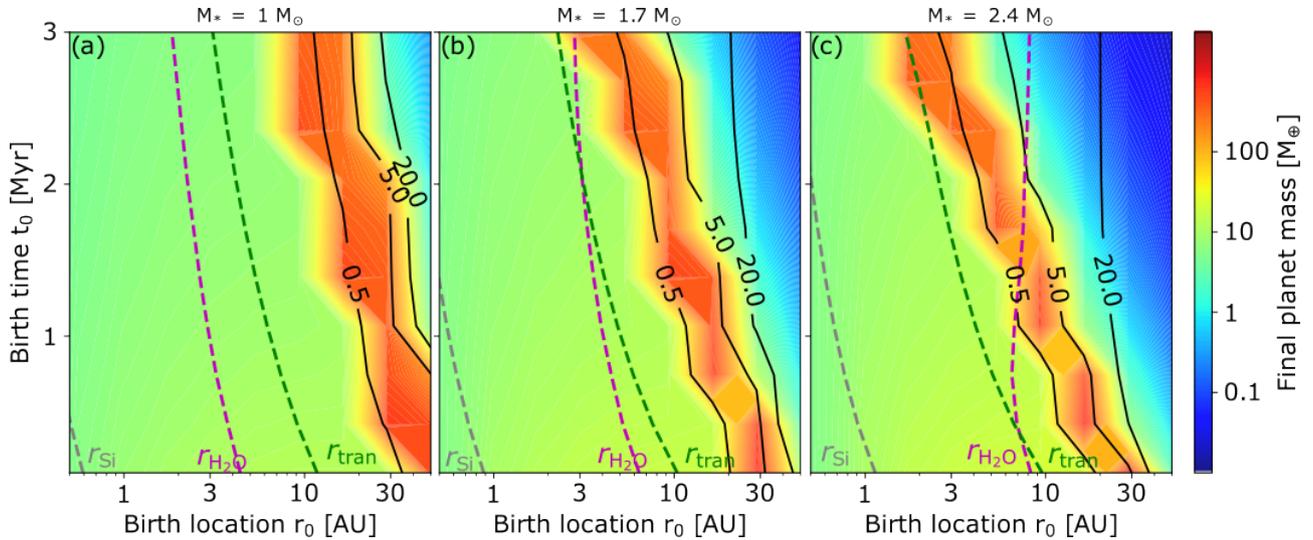

**Figure 7.** Maps for the growth and migration of planets as functions of initial time $t_0$ and $r_0$ around stars of (a) 1 $M_\odot$, (b) 1.7 $M_\odot$, (c) and 2.4 $M_\odot$. The black solid line refers to the final location of the planet, whereas the transition radius, silicate and water sublimation line are marked as green, grey and magenta dashed lines, respectively. The model parameters can be seen in *run-tau* of Table 2.

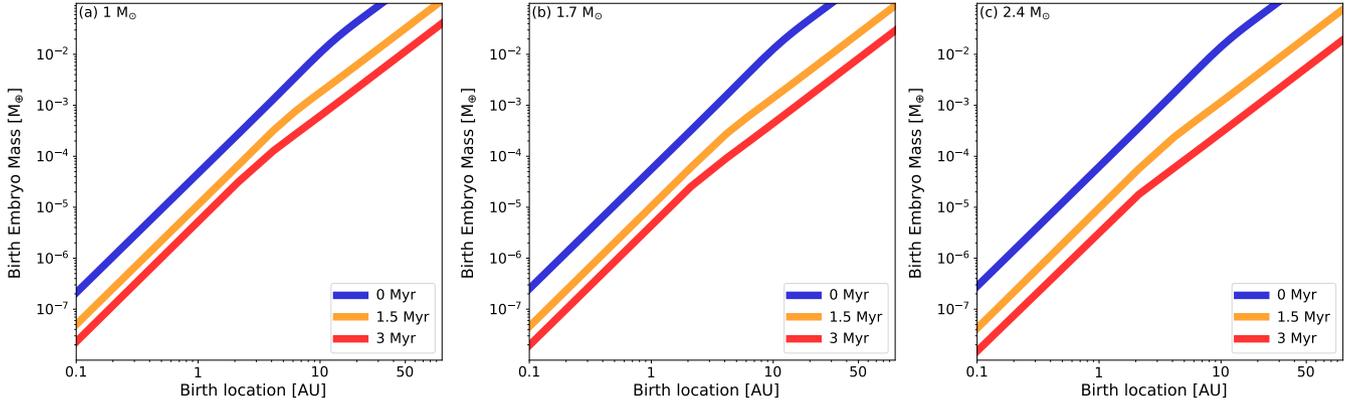

**Figure 8.** Mass of embryo formed by streaming instability as a function of radial distance around stars of 1 $M_\odot$, 1.7 $M_\odot$, and 2.4 $M_\odot$, respectively. The birth time of the embryo is assumed at $t$=0 Myr (blue), 1.5 Myr (orange) and 3 Myr (red). The birth mass of the embryo is higher at earlier time, larger orbital distance and/or around a more massive star.

The self-gravity parameter $\gamma$ that quantifies the relative strength between the gravity and tidal shear is given by

$$\gamma \equiv \frac{4\pi G \rho_g}{\Omega_K^2}. \quad (20)$$

Figure 8 gives the embryo's birth mass as a function of time and radial distance at $M_\star$=1 $M_\odot$, 1.7 $M_\odot$, and 2.4 $M_\odot$. Basically, only embryos formed at $r_0$>10 AU have the masses higher than $10^{-2}$ $M_\oplus$, while embryos born at close-in orbits ($r_0$<1) AU have very low birth masses of <$10^{-5}$ $M_\oplus$.

The migration and growth map is illustrated in Figure 9. The red giant planet formation zone is more horizontally inclined in this case compared to that in Figure 6. This is due to the fact that the embryos have higher initial masses at further out disk locations, especially at early times (Figure 8). These wide-orbit embryos have a much larger initial mass than embryos described in Figure 6 and are able to accrete more material at earlier stages of disk evolution when the accretion rate is highest (Figure 1). Owing to their high masses, such embryos are now able to grow into massive giant planets with orbits of a few AUs.

Furthermore, a new giant planet formation region has appeared in the left corner of Figure 9c. In Figure 6 the embryos born with $10^{-2}$ $M_\oplus$ grow relatively fast at close-in orbits of $r_0 \lesssim 1$ AU. They undergo substantial inward migration when their masses exceed 1 $M_\oplus$. Since their growth fails to compete with fast migration, these embryos quickly reach the inner edge of the disk where both their growth and migration is halted. Nevertheless, in our new study here the embryos at close-in orbits have much lower initial masses. So they take longer time to accrete, and therefore, the onset of efficient inward migration occur at a later time when the disk becomes less massive. Their migration is not as fast as the fiducial run. As such, the embryos can avoid quickly entering the disk edge and grow into giant planets.

Notably, for embryos born early at the 1–10 AU disk region, they still grow relatively fast and migrate interior to the silicate sublimation lines ($r_{si}$) before initiating runaway gas accretion. Our $r_{si}$ (grey) is found very close to the edge of the inner disk ($r_{in}$). Silicate sublimatation occurs at disk temperatures $\geq$ 1500 K, *i.e.* within $r_0$<1.5 AU of our host stars ($M_\star$=1 $M_\odot$, 1.7 $M_\odot$, 2.4 $M_\odot$). In such a circumstance, these embryos finally reach immediate-mass planets of a few tens of Earths.

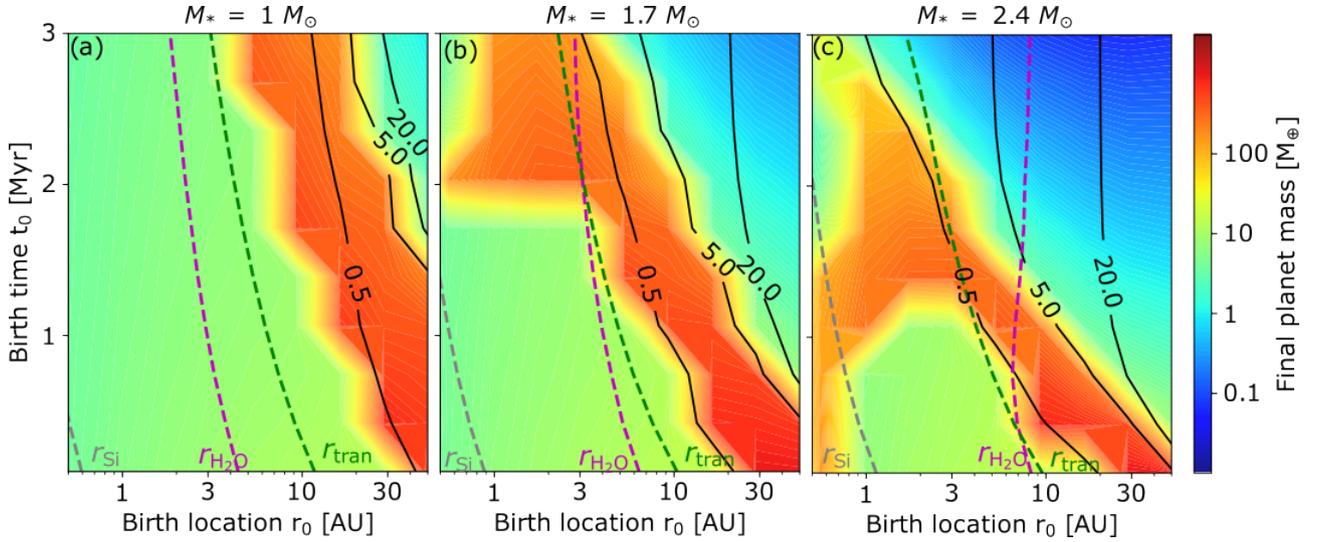

**Figure 9.** Maps for growth and migration of planets with initial masses of embryos computed, rather than fixed. Panels correspond to stars of 1 $M_\odot$ (panel a); 1.7 $M_\odot$ (panel b); and 2.4 $M_\odot$ (panel c), respectively. The black solid line refers to the final location of the planet, whereas the transition radius, silicate, and water sublimation line are marked as green, grey, and magenta dashed lines, respectively.

We also see another distinct shift of the giant planet forming zone between Figure 9a, b, and c as we increase $M_\star$. In Figure 9a, around a 1 $M_\odot$ star, the giant planet forming region is incredibly broad and spans large radial distances for the majority of the 3 Myr. The most massive planet formation ($M_p \geq 1,000 M_\oplus$) occurring between 15-50 AU during the initial 1 Myr. The edge of the giant planet forming zone does move inward with time and by $t_0=3$ Myr, the giant forming region spans 5−25 AU. Around a 1.7 $M_\odot$ star in Figure 9b, we see that while the most massive gas giants still form at large distances in the first 1 Myr is coupled with massive planets formed within the inner disk at times later than 2 Myr. Finally, in Figure 9c around a 2.4 $M_\odot$ star we can observe that giant planet formation is constrained by both birth location and birth time. The giant planet forming region now extends to just 20−50 AU when $t \sim 0$ Myr and planet formation as a whole is less effective at later times and has begun to fail by 3 Myr. The most massive planets still form in the outer regions of the disk but now only for the first $\sim 0.5$ Myr, emphasising the limited ability for embryos to do so.

To summarize, giant planet formation - and planet formation as a whole - is most widely effective around 1 $M_\odot$ and 1.7 $M_\odot$ host stars and while giant planets can form around a 2.4 $M_\odot$ star, they are restricted in where/when they can do so. Lastly, the giant planets are more likely to form in the circumstances when the embryos are born at early times and at moderate radial disk locations, the pebbles are approximately millimeter-sized (or $\tau_s \sim 0.01$), the initial disk sizes are larger and/or the disk accretion rates are higher.

## 5 POPULATION SYNTHESIS STUDY

### 5.1 Description of the model setup

We carry out a population synthesis study, resulting in the final masses and locations of planets formed in our models. This allows us to explore the frequency and efficiency of planet formation with varying initial disk conditions, and to investigate the birth locations corresponding to the resulting planet orbits. The disk model is as described in Section 2.1. We proceed by conducting 1,000 numerical simulations for the growth of a single embryo by Monte Carlo uniform sampling of the initial conditions: $R_{d0}$ sampled between 20 and 200 AU, $r_0$ sampled logarithmically between 0.1 and 100 AU and $t_0$ sampled between 0.1 and 3 Myr. The adopted distributions of disk and stellar parameters are given in Table 3. We perform three classes of simulations defined by the disk pebble's sizes $R_{\rm peb}$, by the birth masses of embryos $M_{p0}$, and the stellar mass $M_\star$, as shown in Table 3.

| Parameter | Description |
|---|---|
| $M_\star$ [ $M_\odot$ ] | 1, 1.7, 2.4 |
| $R_{\rm peb}$ | 1 mm, 100 $\mu m$ |
| $M_{p0}$ [ $M_\oplus$ ] | Eq. (19), $10^{-2}$ |

**Table 3.** Adopted parameter distributions for the population synthesis study in Sect. 5.

### 5.2 Final planetary masses

The resulting planet population of masses vs semimajor axes are illustrated in Figure 10, where the color refers to the final planetary mass of the embryos (low- ($<1 M_\oplus$), intermediate- ($1-299 M_\oplus$), giant- ($>300 M_\oplus$)) and birth location of the embryo, respectively. Note that 300 $M_\oplus$ was selected as the lower mass limit distinguishing giant planets in order to be reflective of the RV survey used to calculate the giant planet occurrence rates in the Reffert et al. (2015) study.

Figure 10 shows that the giant planets in our models generally become less frequent with increasing stellar mass. The planets that have undergone runaway gas accretion in our models are clustered above approximately 10 $M_\oplus$ and within a few AU of the host star due to their effective migration. The occurrence rates we obtain, depicted in Figure 11, decrease by about a factor of two across our stellar mass range for the simulations which take $R_{\rm peb}=1$ mm. In case of $R_{\rm peb}=100 \mu m$, the resulting giant planet numbers are too low to draw meaningful comparisons, as the giant planet formation efficiency is overall very low, below 1%.

Overall in our simulations, the giant planets have occurrences two orders of magnitude lower than any other planets, demonstrating how challenging they are to form. This emphasises that the focus on giant planets in population synthesis studies is a valuable tool to effectively constrain planet formation process overall. The trends we see are the result of a combination of a number of assumptions, effects of which may sometimes be counterintuitive, hence running these models is necessary. If we are to single out one assumption that favours the trend we see in decreasing giant planet formation with stellar mass, this would be our static assumption that the X-ray luminosity scaling with stellar mass remains constant over time, and this implies that disc lifetimes are shorter with increasing stellar mass. However, we know that more massive stars eventually become radiative, producing less X-rays, and their dominant mode of photevaporation shifts to EUV/FUV driven processes (Kunitomo et al. 2021) hence the disc lifetimes may be different.

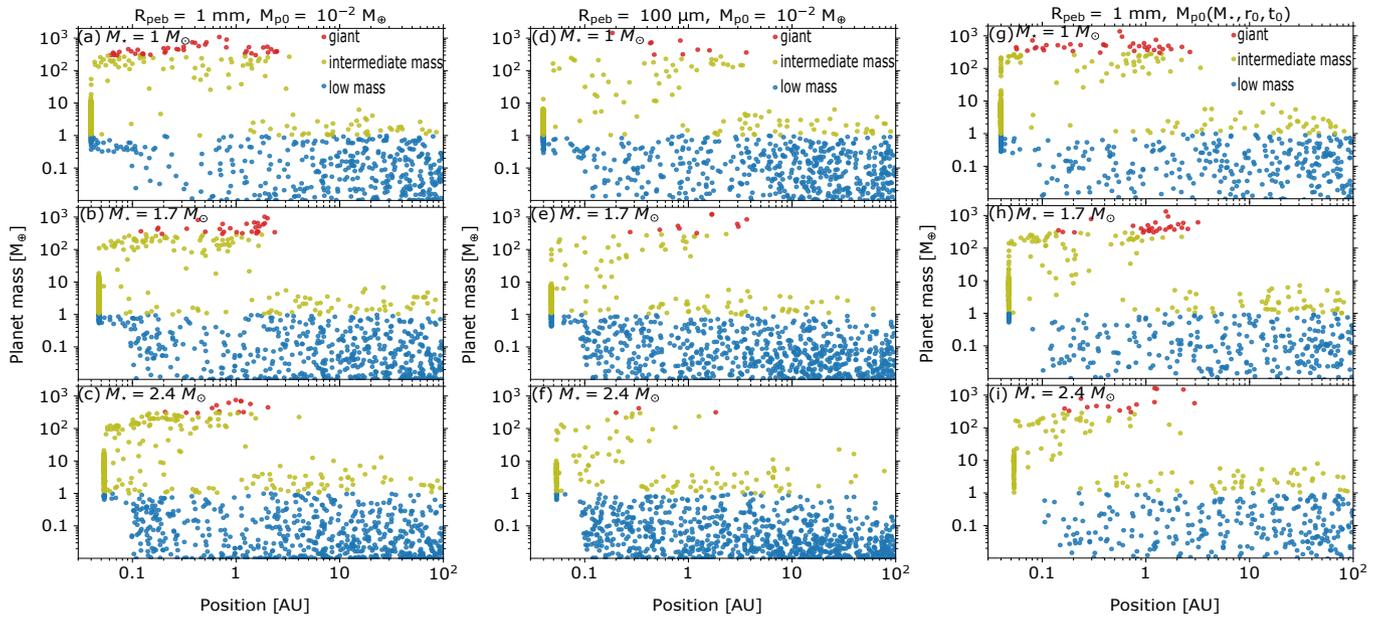

**Figure 10.** Monte Carlo sampling plot of the final mass vs the semi-major axis for single planet systems around stars of $M_\star = 1\,M_\odot$ (a, d, g), $1.7\,M_\odot$ (b, e, h), and $2.4\,M_\odot$ (c, f, i). The color refers to the final planetary mass of the embryo as follows: blue $< 1\,M_\oplus$, green $1-300\,M_\oplus$, and red $> 300\,M_\oplus$. Parameters can be found in Table 3, where $R_{\rm peb}$=1 mm, $M_{\rm p0}$=$10^{-2}\,M_\oplus$ in the left panel (a, b, c), $R_{\rm peb}$=100 $\mu m$ $M_{\rm p0}$=$10^{-2}\,M_\oplus$ in the middle panel (d, e, f), and $R_{\rm peb}$=1 mm, $M_{\rm p0}$ is adopted from Equation (19) in the right panel (g, h, i), respectively. Both gas giants and super-Earths are more numerous around stars of 1 $M_\odot$ than those of 1.7 $M_\odot$ or 2.4 $M_\odot$.

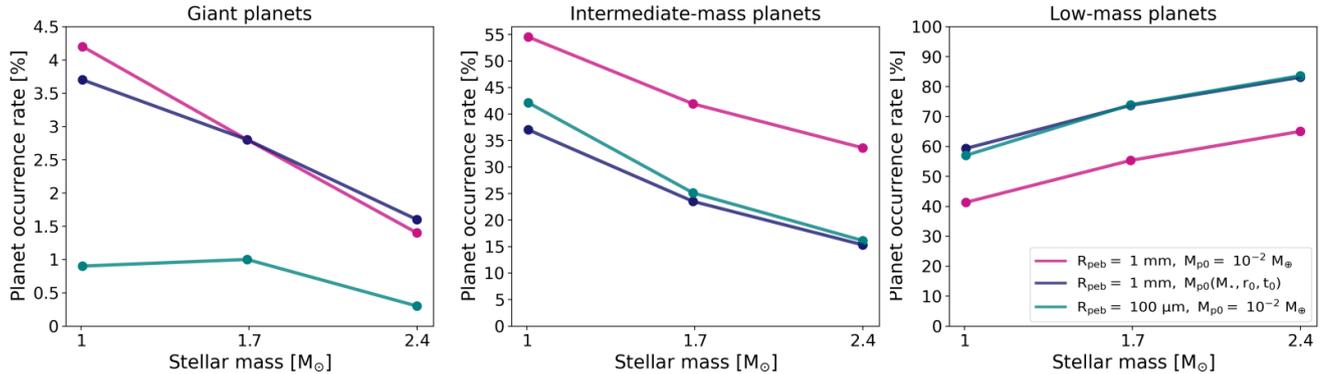

**Figure 11.** Final fractions of planets from Figure 10 in each final mass category (giant ($M_{\rm p} \geq 300\,M_\oplus$, a); intermediate- ($1-299\,M_\oplus$, b); and low- ($< 1\,M_\oplus$, c). The colours refer to the class of simulation: $R_{\rm peb}$=1 mm, $M_{\rm p0}$=$10^{-2}$ in pink; $R_{\rm peb}$=1 mm, $M_{\rm p0}$ in dark blue; and $R_{\rm peb}$=100 $\mu$m, $M_{\rm p0}$=$10^{-2}\,M_\oplus$ in teal. There is a clear trend in our models of the number of giant and intermediate-mass planets decreasing with increasing $M_\odot$.

The simulations with fixed and variable embryo mass, denoted as $M_{\rm p0}$=$10^{-2}\,M_\oplus$ and $M_{\rm p0}(M_\star, r_0, t_0)$, respectively, share similar occurrences of giant planets, whileas the smaller pebble size $R_{\rm peb}$=100 $\mu$m simulation has drastically lower occurrences. These results show that pebble size and stellar mass have more significant impact upon giant planet occurrence rather than the assumptions made on the initial embryo mass. However, in the middle panel of Figure 11, the embryo mass begins to play a role in driving the occurrence rate of intermediate-mass planets ($M_{\rm p}$=$1-300\,M_\oplus$), with the highest occurrence rates corresponding to the fixed embryo mass of $10^{-2}\,M_\oplus$ and $R_{\rm peb}$=1 mm. In the simulation where the embryo mass is computed by Equation 19, the initial embryo mass is lower than $10^{-2}\,M_\oplus$ in most disk regions except beyond a few 10s of AU (Figure 8). Therefore, the occurrence of intermediate-mass planets is also fewer in $M_{p0}(M_\star, r_0, t_0)$ regime compared to that in $R_{peb}$=100 $\mu$m regime (Figure 10). The giant planet occurrence does not have this dependence on initial embryo mass because they form further out in the disk (Figure 9). Overall, the occurrence rates of intermediate-mass planets decrease with increasing stellar mass regardless of the choice of parameters, exhibiting similar behaviour as the occurrence rates of giant planets We can hypothesize that this behaviour is also linked to the shorter disk lifetimes around more massive stars, as assumed in our models.

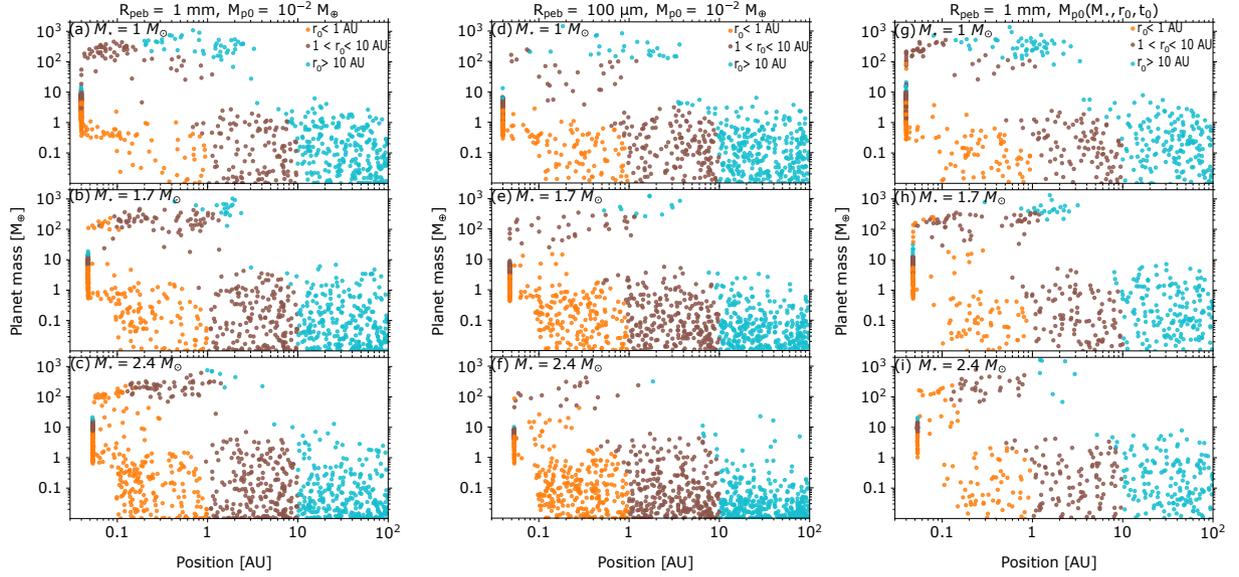

**Figure 12.** Monte Carlo sampling plot of the final mass vs the semi-major axis for single planet systems around stars of $M_\star = 1\,M_\odot$ (a, d, g), $1.7\,M_\odot$ (b, e, h) and $2.4\,M_\odot$ (c, f, i). The color refers to the birth location of the embryo as follows: orange <1 au, brown 1-10 au, and cyan >10 au. Parameters can be found in Table 3, where $R_{peb}$=1 mm, $M_{p0}$=$10^{-2}\,M_\oplus$ in the left panel (a, b, c), $R_{peb}$=100 $\mu$m, $M_{p0}$=$10^{-2}\,M_\oplus$ in the middle panel (d, e, f), and $R_{peb}$=1 mm, $M_{p0}$ is adopted from Equation (19) in the right panel (g, h, i), respectively.

Lastly, in the right panel of Figure 11 we see the increasing occurrence rate of low-mass planets with increasing stellar mass. Because low-mass planets form fast and precede the formation of any more massive planets, their occurrence rates are not as significantly impacted by the shorter disk lifetimes around more massive stars, but other factors scaling with stellar mass prevail (Section 3).

### 5.3 Planet migration

Figure 12 shows the same population synthesis generated in Figure 10 but the colour now reflects the initial birth location of the embryos. This allows us to connect the final planet orbits to their birth locations and visualise the migration in our models. The giant planets in the class of simulations for 1 $M_\odot$ have formed - on average - at further distances than around more massive stars. At the same time in these simulations, there are no giant planets that originate within 1 AU. Both these results go to show that the scale of migration is larger around solar-mass stars, their giant planets come from afar and any embryos growing close-in get trapped at the edge of the inner disk before they have a chance to become giant. As a result, it is only around 1 $M_\odot$ stars that our models produce hot Jupiters. Our hot Jupiter fraction around 1 $M_\odot$ star is $\sim$ 0.7–0.8% for the $R_{peb}$=1 mm simulations which is in-keeping with current demographic studies that show that hot Jupiters are relatively rare and yield RV occurrence rates of $\sim$ 0.8–1.2 % around Sun-like stars (Mayor et al. 2011; Wright et al. 2012; Wittenmyer et al. 2020).

In comparison the scale of migration around stars of higher mass is progressively smaller. Their giant planet population is dominated by planets that formed between 1–10 AU and migrated to orbits of 0.1–1 AU (brown). As the stellar mass increases, there are no giant planets found within 0.1 AU in our simulations. Instead, planets with final distances <0.1 AU grow to - at most - a few hundred $M_\oplus$s. The majority of such planets have birth locations of <1 AU (orange) and were able to avoid fast inward migration to the edge of the inner disk and instead grow into sub-giants. Migration timescale shown in Equation 17 is directly proportional to stellar mass while the disk lifetime decreases, which explains the trends that we see above.

While the number of giant planets in the middle panel of Figure 12 (d, e, f) is notably reduced, the birth locations of the few successful giant and sub-giant planets share a similar pattern as the $R_{peb}$=1 mm simulations. The reduced number of giant planets can be explained by smaller dust grain size as it takes longer for such embryos acquire enough solid pebbles to exceed their respective isolation masses and undergo runaway gas accretion. Thus, embryos in this setup born at $r_0 > 10$ AU are at an advantage as they must migrate for a longer timescale allowing them to become massive enough. No planets initially located at $r_0 < 1$ AU are able to grow larger than a few $M_\oplus$s.

### 5.4 Model and Comparison to Observations

From our fraction of giant planets presented in Figure 11, we find a consistent decrease in the number of giant planets with increasing $M_\star$. This result is at odds with the observational surveys that report a peak in the giant planet occurrence rate between 1.7−2 $M_\odot$ (Reffert et al. 2015; Wolthoff et al. 2022). A few examples of how to improve our current models are as follows:

(i) Employing an evolving pre-main sequence luminosity to better reflect parameters such as changing midplane disk temperatures and efficient radial drift (Miley et al. 2020; Pinilla et al. 2022). Temper-ature plays an indirect role in setting the disk scale height and local sound speed, both parameters being involved in processes of pebble accretion and planet migration, and both clearly central to our results. It would be important to then examine the realistic models and their impact on planet formation.

(ii) Introducing a stellar mass dependence of the mass accretion rate. As demonstrated in Figures 3 and 5, giant planet formation is more efficient at higher accretion rates, and the correlation of accretion rate with stellar mass, seen observationally paper too (Manara et al. 2017; Wichittanakom et al. 2020; Iglesias et al. 2023), would hinder giant planet formation around a 1 $M_\odot$ which in current models is too efficient.

(iii) Introducing a stellar mass dependence of the metallicity. In Section 3, we show that metallicity is an important parameter in forming giant planets, and indeed observations of discs show higher solid (dust) masses in discs around more massive stars (Pascucci et al. 2016) Thus, we could steepen the dependence of metallicity on stellar mass to boost giant planet formation around higher mass stars.

Due to the complexity of the model and interdependencies of various processes, we cannot speculate how these improvements would impact planet occurrences beyond simplistic explorations in Section 3 but are clearly motivated improvements that would make our model more realistic. Proper simulations that include these improvements will be the subject of a future paper.

We find that our hot Jupiter occurrence rate around 1 $M_\odot$ star is ∼ 0.7−0.8% (Figure 12 a) and d) is in agreement with RV surveys around Sun-like stars (Mayor et al. 2011; Wright et al. 2012; Wittenmyer et al. 2020).Additionally, we do not find any hot Jupiters at distances <0.1 AU (∼ 10 day period) around 1.7 $M_\odot$ or 2.4 $M_\odot$. Beleznay & Kunimoto (2022) found that the abundance of hot Jupiters decreases with increasing stellar mass, reporting an occurrence rate of 0.29 ± 0.05 for A-type stars - consistent with trends seen in main sequence (MS) FGK stars from the Kepler mission (Mulders et al. 2015a; Kunimoto & Matthews 2020; Beleznay & Kunimoto 2022). Similarly, Sabotta et al. (2019) find only one close-in giant planet in their sample of 166 A-type stars from Kepler. Our lack of such hot Jupiters in our 1.7 $M_\odot$ and 2.4 $M_\odot$ simulations agrees with this trend and could be related our assumption that X-ray photoevaporation increases with increasing stellar mass combined with the slower rate of migration (Equation 17). This results in our buildup of sub-giants trapped at the edge of the inner disk discussed in Section 5.3.

## 6 CONCLUSIONS

In this work we employ and extend the pebble-driven core accretion models (Liu et al. 2019, 2020) to study how giant planets form. To this end, we focus on the stellar mass range of 1−2.4 $M_\odot$, where the strongest trends of giant planet occurrence are found. We find that giant planet formation is most favourable in disks with a large initial disk size; a high mass accretion rate; high metallicity (dust-to-gas flux ratio); and pebbles of ∼ 1-mm in size. Even with these idealised conditions, giant planet formation as a whole is incredibly challenging around a 2.4 $M_\odot$ star due to the slow rate of migration combined with fast disk evolution in our model. There is a favoured orbital distance where embryos are able to grow to giant planets due to a combination effect of slow migration and rapid gas disk dissipation (right panels of Figures 4 and 5). This location is 10-30 AU initially, and moves to a few AU over the first three Myr of disc evolution, in general (Figure 6, 7, and 9). Important to note in this context is that the proximity to the water snow line does not seem to play a decisive role in these models (no implementation of enhanced core accretion by icy pebbles that "stick" together more efficiently (Okuzumi et al. 2012; Drazkowska & Alibert 2017; Hyodo et al. 2021)) – more general physical disk conditions are thus much more relevant here.

Using the optimal initial conditions for giant planet formation, we generate growth and migration maps to further explore the trends of giant planet formation. We observe a similar decline of the giant planet occurrence rate with stellar mass at the higher end of stellar masses, as described in Jones et al. (2015); Reffert et al. (2015); Wolthoff et al. (2022) but fail to replicate the observed peak when $M_\star$=1.68−1.9 $M_\odot$. In our models, giant planets are more likely to form around 1 $M_\odot$ stars than around more massive stars - *e.g.* $M_\star$=1.7, 2.4 $M_\odot$ - (Figures 6, 7, 9, and 10. This indicates that some parameters - *e.g.* mass accretion rate; pre-main sequence luminosity evolution, metallicity - need to be employed with more realistic $M_\star$-dependencies in order to achieve the observed rise and fall of the giant planet occurrence rate. Planet migration is less pronounced in disks around 2.4 $M_\odot$ than around 1 $M_\odot$ and 1.7 $M_\odot$ stars. Therefore, in such systems, embryos can avoid fast inward migration and reaching the inner edge of the disk and running out of gaseous material necessary to accrete/migrate, so they can grow into close-in giant planets (Figure 9 and 12). The code allowed us to follow the water content through the formation process and find that the water fraction accreted during the formation of giant planets decreases with $M_\star$ (Figure A1).

## ACKNOWLEDGEMENTS

The authors appreciate the thoughtful comments of the referee. We also thank Sabine Reffert, Richard Booth, Daniela Iglesias, and Dimitri Veras for their useful discussions and comments. HFJ is funded by the Science and Technology Facilities Council (Project Reference: 2441776). OP acknowledges support from the Science and Technology Facilities Council, grant number ST/T000287/1. BL is supported by National Natural Science Foundation of China (Nos. 12222303, 12173035 and 12111530175), the start-up grant of the Bairen program from Zhejiang University and the Fundamental Research Funds for the Central Universities (2022-KYY-506107-0001,226-2022-00216).

## DATA AVAILABILITY

The data underlying this article will be shared on reasonable requests to the corresponding author.

## APPENDIX A: WATER CONTENT

We assume that the pebbles beyond $r_{H_2O}$ consist of 35% water ice and 65% silicates ("wet" pebbles), whereas interior to $r_{H_2O}$ the water ice is sublimated and pebbles become purely rocky ("dry" pebbles). The locations where embryos accrete pebbles are responsible for the water content in the planetary cores, which is a crucial indicator of their formation locations and migration dynamics.

In Figure 6, the final planetary cores would become either water-deficit or water-rich when the embryos at grow $r_0 \ll r_{H_2O}$ or $r_0 \gg r_{H_2O}$. This is expected since the embryos in these regions would accrete either "dry" or "wet" pebbles during their lifetime. Only embryos formed close to $r_{H_2O}$ would contain moderate water fraction, ranging from a few percent to ~20%. The exact water content is a competition between migration (relative movement between the planet and water ice line) and planet growth.

Combing Figure 6 a and A1 we find that the cores of giant planets contain substantial water in systems around stars of 1 $M_\odot$. Noticeably, the water fraction of giant planet cores decrease as the masses of their stellar hosts increase (Figure 6 and A1 b and c). The cores of giant planets formed after 1 Myr in systems around stars of 3 $M_\odot$ only accrete dry pebbles and become water-deficit. Again, the reason for this is because the migration rate decreases with stellar mass, while the growth rate is modestly dependent on stellar mass.

Our model predicts the water in the giant planetary cores decreases with the masses of their central hosts. This water as well as other volatile heavy elements in the cores might be diluted into their atmosphere, illustrating a signature of super-stellar metallicity. In future, examining the composition of giant plants around stars of various masses by James Webb Space Telescope could be a valuable probe for our model.

This paper has been typeset from a TEX/LATEX file prepared by the author.

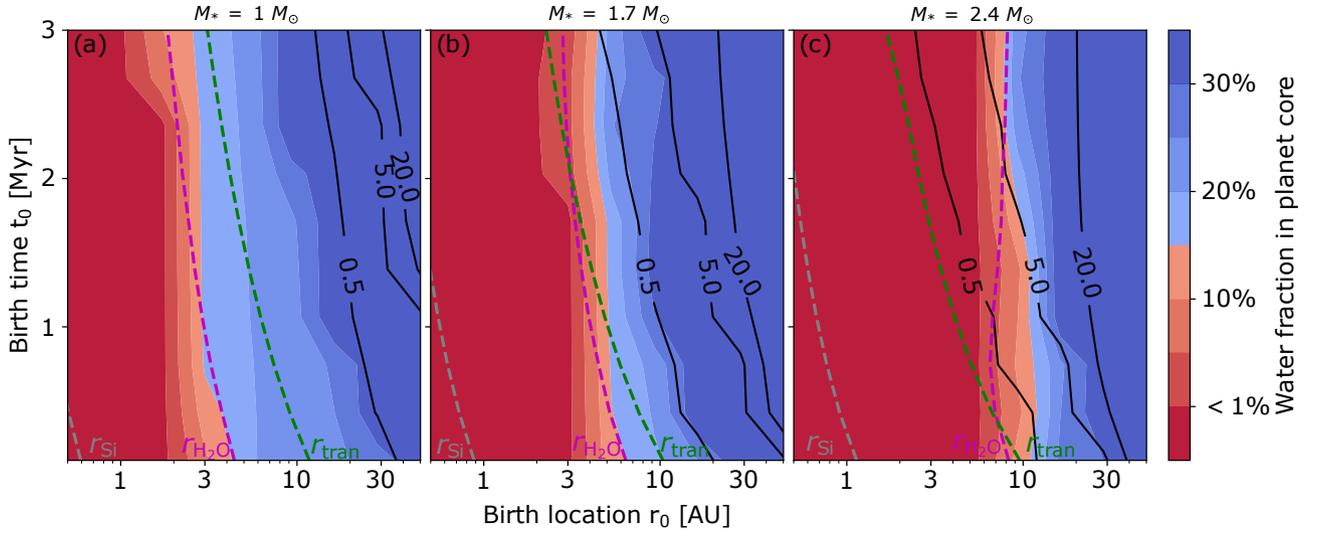

**Figure A1.** Maps for the water content of planets as functions of initial time $t_0$ and location $r_0$ around stars of 1 $M_\odot$ (left) and 1.7 $M_\odot$ (middle) and 2.4 $M_\odot$ (right), respectively. The black solid line refers to the final location of the planet, whereas the transition radius, silicate and water sublimation line are marked as green, grey and magenta dashed lines, respectively. The model parameters can be seen in *run-fid* of Table 2. The water fraction in planetary cores ranges from a few percent to ∼20% when the embryos are born near the water ice line.